\documentclass[11pt,a4paper]{article}
\pdfoutput=1

\usepackage[T1]{fontenc}
\usepackage{graphicx, color, subfig, caption, a4wide, fancyvrb, upquote}
\usepackage{booktabs, hyperref}
\usepackage{longtable, multirow, enumitem}
\usepackage{amsmath, amsfonts}
\usepackage{chicago}

\usepackage[round]{natbib}
\usepackage[margin=1in]{geometry}

\usepackage[scaled=0.9]{inconsolata}  
\DefineVerbatimEnvironment{example}{Verbatim}{baselinestretch=0.9}

\makeatletter
\newcommand\code{\bgroup\@makeother\_\@makeother\~\@makeother\$\@codex}
\def\@codex#1{{\normalfont\ttfamily\hyphenchar\font=-1 #1}\egroup}
\makeatother

\newcommand{\R}{R}
\newcommand{\pkg}[1]{{\fontseries{b}\selectfont #1}}
\newcommand{\CRANpkg}[1]{\href{http://CRAN.R-project.org/package=#1}{\pkg{#1}}}%

\newcommand{\BIC}{\mathrm{BIC}}

\newcommand{\loglik}{\ell}
\renewcommand{\hat}{\widehat}

\definecolor{dodgerblue2}{RGB}{28,134,238}
\definecolor{red3}{RGB}{205,0,0}

\begin{document}

\title{On some extensions to \pkg{GA} package:\\hybrid optimisation, parallelisation and islands evolution}
\author{%
Luca Scrucca \\ 
Department of Economics \\ 
Universit\`a degli Studi di Perugia\\
\texttt{luca.scrucca@unipg.it}
}
\date{\today}
\maketitle

\begin{abstract}
Genetic algorithms are stochastic iterative algorithms in which a population of individuals evolve by emulating the process of biological evolution and natural selection.
The \R\ package \pkg{GA} provides a collection of general purpose functions for optimisation using genetic algorithms. 
This paper describes some enhancements recently introduced in version 3 of the package. 
In particular, hybrid GAs have been implemented by including the option to perform local searches during the evolution. This allows to combine the power of genetic algorithms with the speed of a local optimiser. 
Another major improvement is the provision of facilities for parallel computing. Parallelisation has been implemented using both the master-slave approach and the islands evolution model.
Several examples of usage are presented, with both real-world data examples and benchmark functions, showing that often high-quality solutions can be obtained more efficiently.\\

\noindent {\it Keywords:} Genetic algorithms, Evolutionary computing, Hybrid algorithms, Parallel computing, R, GA package.
\end{abstract}



\section{Introduction}

Optimisation problems of both practical and theoretical importance deal with the search of an optimal configuration for a set of variables to achieve some specified goals. 
Potential solutions may be encoded with real-valued, discrete, binary or permutation decision variables depending on the problem to be solved. 
Optimisation methods for real-valued functions can be roughly classified into two groups: direct and gradient-based methods 
(\citealp{Chong:Zak:2013}; \citealp[Chap. 2]{Givens:Hoeting:2013}).
In direct search methods only the objective function is used to guide the search strategy, whereas gradient-based methods consider the first and/or second-order derivatives of the objective function during the search process. 
Constraints may be present and are usually taken into account in the definition of the objective function or in the decision variables representation.
Direct search methods can be applied without modifications to many optimisation tasks, but they are usually slow requiring many function evaluations for convergence. 
On the contrary, gradient-based methods quickly converge to an optimal solution, but are not efficient in non-differentiable or discontinuous problems. 
Both direct and gradient-based techniques depend on the chosen initial starting values, so they can get stuck in suboptimal solutions. 
Furthermore, they are not efficient in handling problems with discrete decision variables, and cannot be efficiently implemented on parallel machines.
Problems where the decision variables are expressed using discrete or binary values are usually referred to as combinatorial optimisation problems, and consist in searching for the best solution from a set of discrete items (\citealp{Papadimitriou:Steiglitz:1998}; \citealp[Chap. 3]{Givens:Hoeting:2013}).
Typical examples are the knapsack problem, the minimum spanning tree, the traveling salesman problem, and the vehicle routing problem. Although in principle these type of problems can be solved with exact algorithms, the time required to solve them increases exponentially as the size of the problem grows. 

A large number of heuristics and metaheuristics algorithms have been proposed for solving complex optimisation tasks . 
Specific (ad-hoc) heuristic techniques are able to identify solutions in a reasonably short amount of time, but the solutions obtained are generally not guaranteed to be optimal or accurate.
On the contrary, metaheuristics offer a tradeoff between exact and heuristics methods, in the sense that they are generic techniques that offer good solutions, often the global optimal value sought, in a moderate execution time by efficiently and effectively exploring the search space \citep{Luke:2013}.
This class of algorithms typically implements some form of stochastic optimisation and includes:
Evolutionary Algorithm \citep[EA;][]{Back:Fogel:Michalewicz:2000:vol1, Back:Fogel:Michalewicz:2000:vol2}, 
Iterated Local Search \citep[ILS;][]{Lourenco:Stutzle:2003},
Simulated Annealing \citep[SA;][]{Kirkpatrick:Gelatt:Vecchi:1983},
Tabu Search \citep[TS;][]{Glover:Laguna:2013},
and Ant Colony Optimisation \citep[ACO;][]{Dorigo:Stutzle:2004}.

EAs are stochastic iterative algorithms in which a population of individuals evolve by emulating the biological processes observed in natural evolution and genetics \citep{Eiben:Smith:2003, DeJong:2006, Simon:2013}.
Each individual of the population represents a tentative solution to the problem. The quality of the proposed solution is expressed by the value of a fitness function assigned to each individual. This value is then used by EAs to guide the search and improve the fitness of the population.
Compared to other metaheuristics algorithms, EAs are able to balance between exploration of new areas of the search space and exploitation of good solutions. 
The trade-off between exploration and exploitation is controlled by some tuning parameters, such as the population size, the genetics operators (i.e. selection, crossover, and mutation), and the probability of applying them.
Genetic Algorithms (GAs) are search and optimisation procedures that are motivated by the principles of natural genetics and natural selection. 
GAs are the ``earliest, most well-known, and most widely-used EAs'' \citep[p. 35]{Simon:2013}.

\R\ offers several tools for solving optimisation problems. A comprehensive listing of available packages is contained in the CRAN task view on ``Optimization and Mathematical Programming'' \citep{CRANTaskView:Optimization}. 
An extensive treatment of optimisation techniques applied to problems that arise in statistics and how to solve them using \R\ is provided by \citet{Nash:2014}. A gentle introduction to metaheuristics optimisation methods in \R\ is contained in \citet{Cortez:2014}. 
The \R\ package \CRANpkg{GA} is a flexible general-purpose set of tools for optimisation using genetic algorithms and it is fully described in \citet{Scrucca:2013}. Real-valued, integer, binary and permutation GAs are implemented, whether constrained or not. Users can easily define their own objective function depending on the problem at hand. Several genetic operators for selection, crossover, and mutation are available, and more can be defined by experienced \R\ users. 

This paper describes some recent additions to the \pkg{GA} package. The first improvement involves the option to use hybrid GAs. Although GAs are able to identify the region of the search space where the global optimum is located, they are not especially fast at finding the optimum when in a locally quadratic region. 
Hybrid GAs combine the power of GAs with the speed of a local optimiser, allowing researchers to find a global solution more efficiently than with the conventional evolutionary algorithms.
Because GAs can be easily and conveniently executed in parallel machines, the second area of improvement is that associated with parallel computing. Two approaches, the master-slave and islands models, have been implemented and are fully described. 
Several examples, using both real-world data examples and benchmark functions, are presented and discussed.

\section{GA package}

In the following we assume that the reader has already installed the latest version ($\ge 3.0$) of the package from CRAN with
\begin{example}
> install.packages("GA")
\end{example}
and the package is loaded into an \R\ session using the usual command
\begin{example}
> library(GA)
\end{example}

\section{Hybrid genetic algorithms}
\label{sec:hga}

EAs are very good at identifying near-optimal regions of the search space (\textit{exploration}), but they can take a relatively long time to locate the exact local optimum in the region of interest (\textit{exploitation}). 
More effective algorithms might try to incorporate efficient local search algorithms into EAs.  
There are different ways in which local searches or problem-specific information can be integrated in EAs \citep[see ][Chap. 10]{Eiben:Smith:2003}. 
For instance, a local search may be started from the best solution found by a GA after a certain number of iterations, so that, once a promising region is identified, the convergence to the global optimum can be speed up. 

These evolutionary methods have been named in various ways, such as \emph{hybrid GAs}, \emph{memetic GAs}, and \emph{genetic local search algorithms}. 
Some have argued that the inclusion of a local search in GAs implies the use of a form of Lamarckian evolution. This fact has been criticised from a biological point of view, but ``despite the theoretical objections, hybrid genetic algorithms typically do well at optimization tasks'' \citep[p. 82]{Whitley:1994}.

In case of real-valued optimisation problems, the \pkg{GA} package provides a simple to use implementation of hybrid GAs by setting the argument \code{optim = TRUE} in a \code{ga()} function call. 
This allows to perform local searches using the base \R\ function \code{optim()}, which makes available general-purpose optimisation methods, such as Nelder–Mead, quasi-Newton with and without box constraints, and conjugate-gradient algorithms. 

Having set \code{optim = TRUE}, the local search method to be used and other parameters can be controlled with the optional argument \code{optimArgs}. This must be a list with the following structure and defaults:
\begin{example}
optimArgs = list(method = "L-BFGS-B", 
                 poptim = 0.05,
                 pressel = 0.5,
                 control = list(fnscale = -1, maxit = 100))
\end{example}
where
\begin{longtable}[l]{lp{0.75\textwidth}}
\code{method} & The method to be used among those available in \code{optim} function (see Details section in \code{help(optim)}). By default, the BFGS with box constraints is used, where the bounds are those provided in the \code{ga()} function call).\\
\code{poptim} & 
A value in the range $(0,1)$ which gives the the probability of applying the local search at each iteration.\\
\code{pressel} & 
A value in the range $(0,1)$ which specifies the pressure selection.\\
\code{control} & 
A list of parameters for fine tuning the \code{optim} algorithm (see \code{help(optim)} for details).
\end{longtable}

In the implementation available in \pkg{GA}, the local search is applied stochastically during the GA iterations with probability \code{poptim} $\in [0,1]$; by default, once every $1/0.05 = 20$ iterations on average. 
The local search algorithm is started from a random selected solution drawn with probability proportional to fitness and with the selection process controlled by the parameter \code{pressel} $\in [0,1]$. 
The latter value is used in the function \code{optimProbsel()} for computing the probability of selection for each individual of the genetic population. Smaller values of \code{pressel} tend to assign equal probabilities to all the solutions, and larger values tend to assign larger values to those solutions having better fitness. 
As an example, consider the following output which presents a vector of fitness values \code{f} assgined to different solutions, and the corresponding probabilities of selection obtained by varying the selection pressure parameter:
\begin{example}
> f <- c(1, 2, 5, 10, 100)
> data.frame(f = f,
             "0"   = optimProbsel(f, 0),
             "0.2" = optimProbsel(f, 0.2),
             "0.5" = optimProbsel(f, 0.5), 
             "0.9" = optimProbsel(f, 0.9),
             "1"   = optimProbsel(f, 1),
             check.names = FALSE)
    f   0    0.2     0.5     0.9         1
1   1 0.2 0.1218 0.03226 0.00009 4.930e-32
2   2 0.2 0.1523 0.06452 0.00090 3.309e-24
3   5 0.2 0.1904 0.12903 0.00900 2.220e-16
4  10 0.2 0.2380 0.25806 0.09000 1.490e-08
5 100 0.2 0.2975 0.51613 0.90001 1.000e+00
\end{example}
When no pressure selection is set, i.e. at 0, the same probability is assigned to all. Larger probabilities are assigned to larger $f$ values as the pressure value increases. In the extreme case of pressure selection equal to 1, only the largest $f$ has assigned a probability of selection equal to 1, whereas the others have no chance of being selected. 

When a \code{ga()} function call is issued with \code{optim = TRUE}, a local search is always applied at the end of GA evolution (even in case of \code{poptim = 0}), but now starting from the solution with the highest fitness value. The rationale for this is to allow for local optimisation as a final improvement step. 

\subsection{Portfolio selection}

In portfolio selection the goal is to find the optimal portfolio, i.e. the portfolio that provides the highest return and lowest risk. This is achieved by choosing the optimal set of proportions of various financial assets \citep[Chap. 16]{Ruppert:Matteson:2015}.
In this section an example of mean–variance efficient portfolio selection \citep[Chap. 13]{Gilli:etal:2011} is illustrated.

Suppose we have selected 10 stocks from which to build a portfolio. We want to determine how much of each stock to include in our portfolio. 
The \emph{expected return rate} of our portfolio is
$$
E(R) = \sum_{i=1}^{10} w_i E(R_i),
$$
where $E(R_i)$ is the expected return rate on asset $i$, and $w_i$ is the fraction of the portfolio value due to asset $i$. Note that the portfolio weights $w_i$ must satisfy the constraints $w_i \ge 0$, and $\sum_{i=1}^{10} w_i = 1$.
At the same time, we want to minimise the \emph{variance of portfolio returns} given by
$$
\sigma^2_p = w' \Sigma w,
$$
where $\Sigma$ is the covariance matrix of stocks returns, and $w' = (w_1, \ldots, w_{10})$, under the constraint that the portfolio must have a minimum expected return of 1\%, i.e $E(R) \ge 0.01$.

\begin{figure}[htbp]
  \centering
  \includegraphics[width=\textwidth]{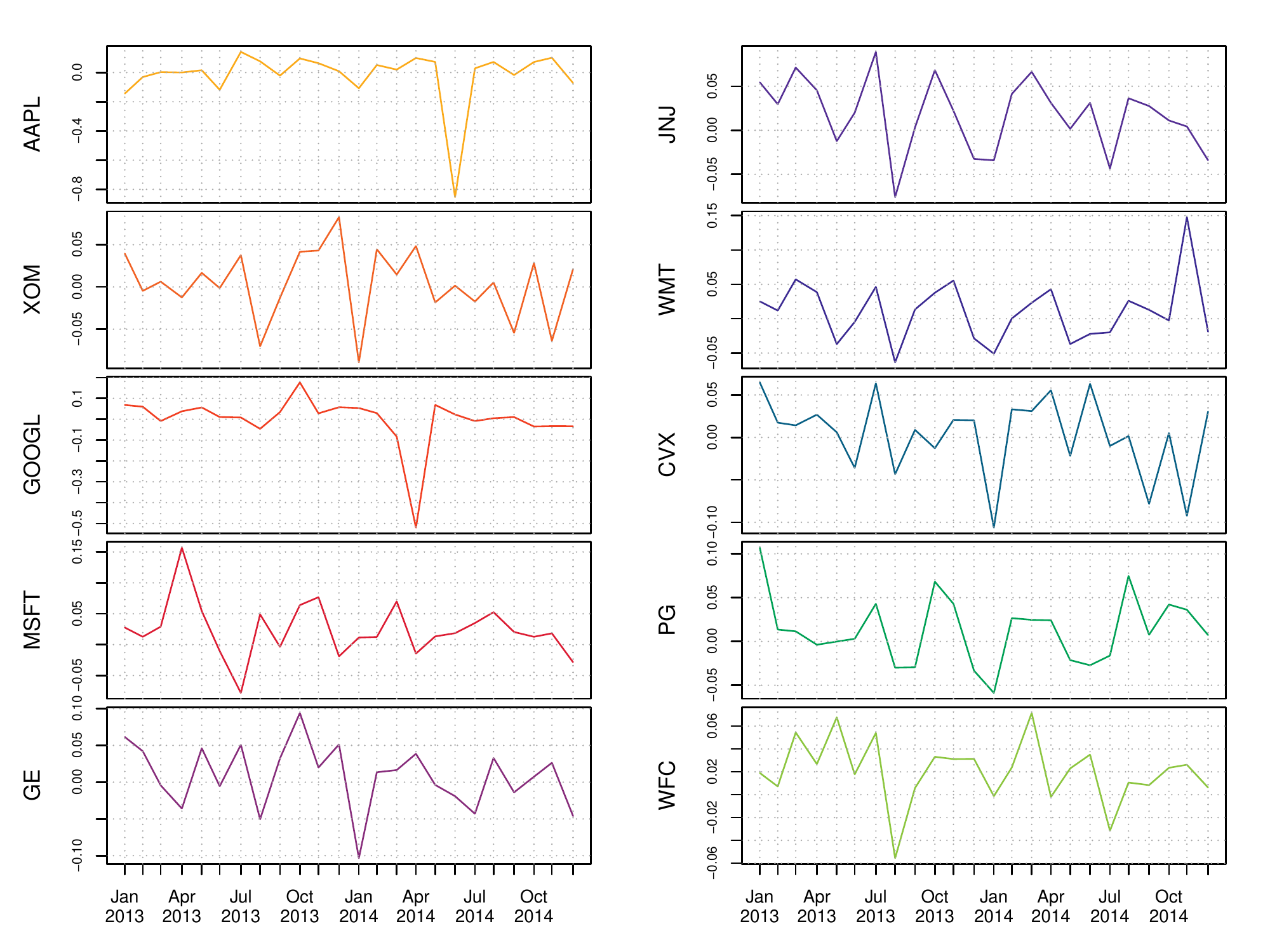}
  \caption{Monthly return rates for a portfolio of selected stocks.}
  \label{fig1:portfolio}
\end{figure}

Consider the following stocks with monthly return rates obtained by Yahoo finance using the \CRANpkg{quantmod} package:
\begin{example}
> library(quantmod)
> myStocks <- c("AAPL", "XOM", "GOOGL", "MSFT", "GE", "JNJ", "WMT", "CVX", "PG", "WFC")
> getSymbols(myStocks, src = "yahoo")
> returns <- lapply(myStocks, function(s) 
                              monthlyReturn(eval(parse(text = s)),
                                            subset = "2013::2014"))
> returns <- do.call(cbind,returns)
> colnames(returns) <- myStocks
\end{example}
The monthly return rates for the portfolio stocks are shown in Figure~\ref{fig1:portfolio} and obtained with the code:
\begin{example}
> library(timeSeries)
> plot(as.timeSeries(returns), at = "chic", minor.ticks="month", 
       mar.multi = c(0.2, 5.1, 0.2, 1.1), oma.multi = c(4, 0, 4, 0),
       col = .colorwheelPalette(10), cex.lab = 0.8, cex.axis = 0.8)
> title("Portfolio Returns")
\end{example}

\begin{figure}[htbp]
\centering\footnotesize
\begin{minipage}{0.4\textwidth}
  \centering
  \includegraphics[width=\textwidth]{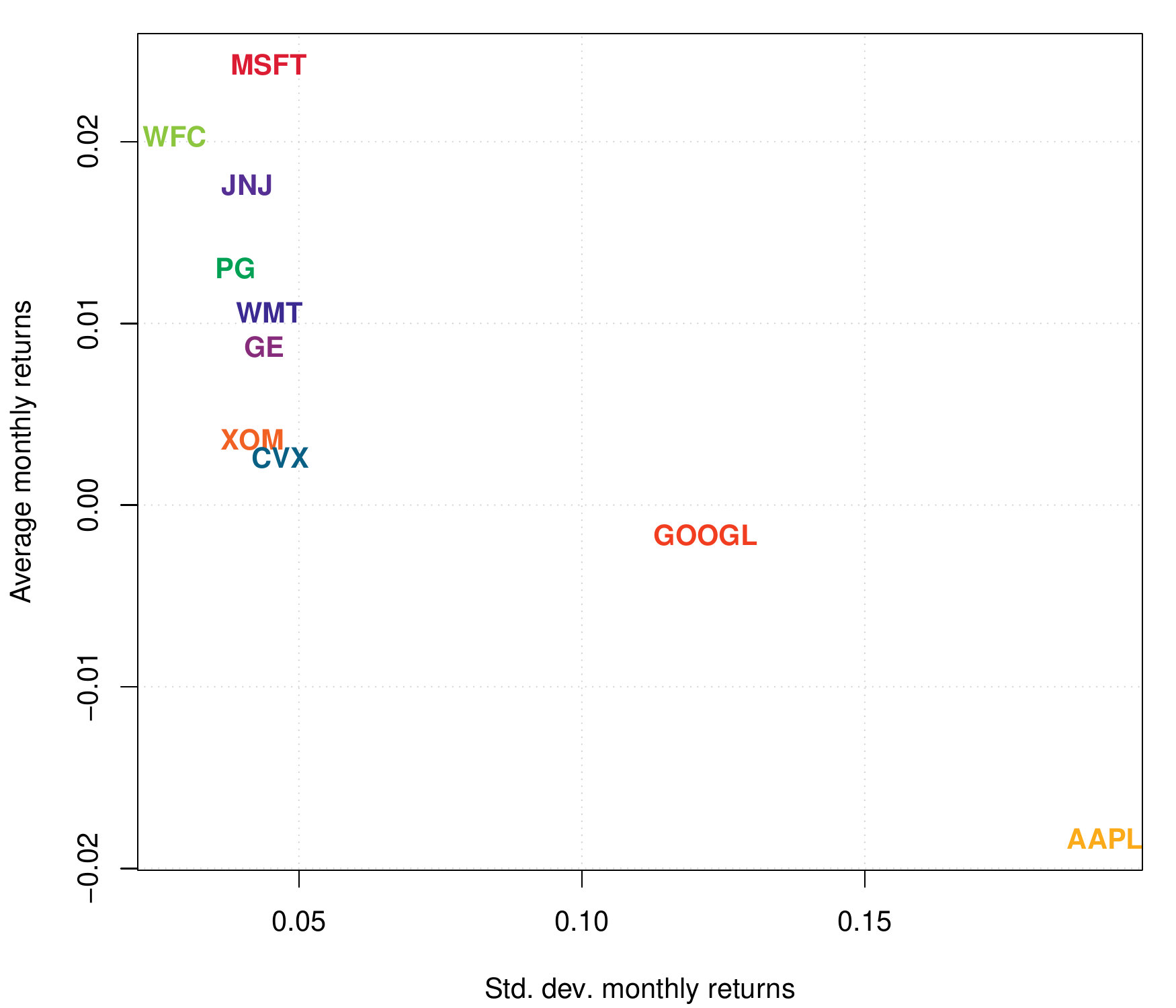}
  (a)
\end{minipage}
\begin{minipage}{0.4\textwidth}
  \centering
  \includegraphics[width=\textwidth]{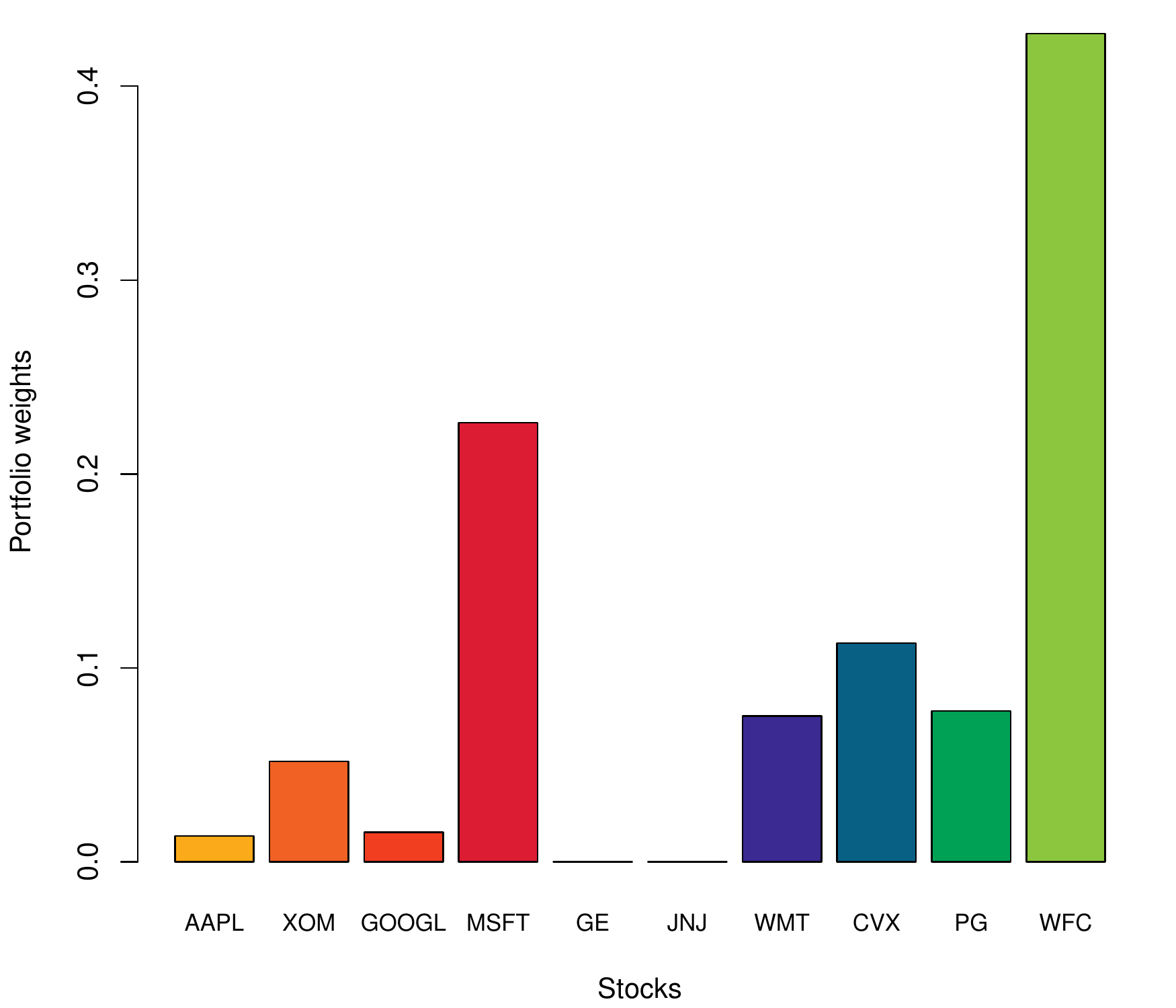}
  (b)
\end{minipage} 
\begin{minipage}{\textwidth}
  \centering
  \includegraphics[width=0.9\textwidth]{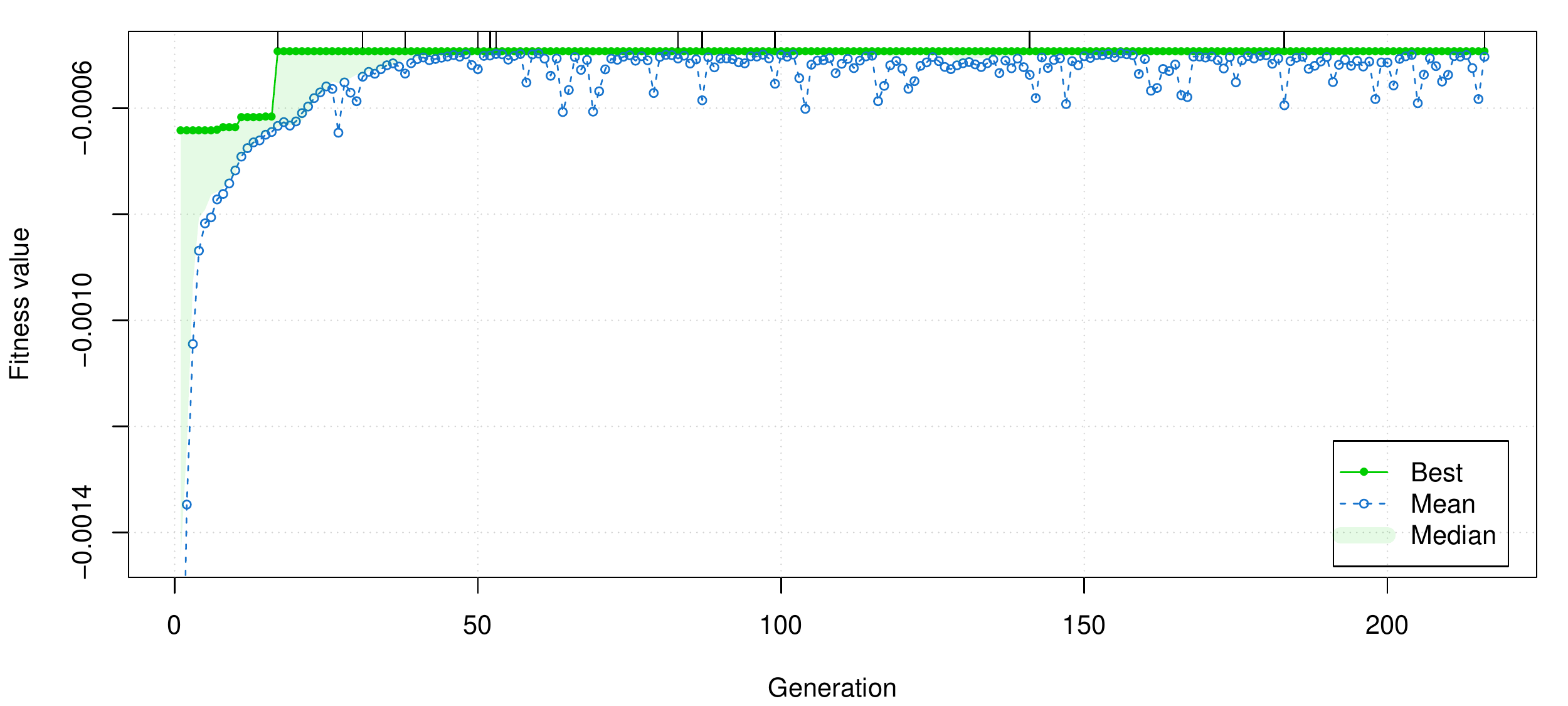}\\
  (c)
\end{minipage} 
  \caption{(a) Plot of average monthly returns vs the standard deviation for the selected stocks. (b) Portfolio stocks composition estimated by HGA. (c) Trace of HGA iterations.}
  \label{fig2-4:portfolio}
\end{figure}

Summary statistics for the portfolio stocks are computed as:
\begin{example}
> nStocks <- ncol(returns) # number of portfolio assets
> R <- colMeans(returns)   # average monthly returns
> S <- cov(returns)        # covariance matrix of monthly returns
> s <- sqrt(diag(S))       # volatility of monthly returns
> plot(s, R, type = "n", panel.first = grid(),
       xlab = "Std. dev. monthly returns", ylab = "Average monthly returns")
> text(s, R, names(R), col = .colorwheelPalette(10), font = 2)
\end{example}
The last two commands draw a graph of the average vs standard deviation for the monthly returns (see Figure~\ref{fig2-4:portfolio}a). From this graph we can see that there exists a high degree of heterogenity among stocks, with AAPL having the largest standard deviation and negative average return, whereas some stocks have small volatility and high returns, such as WFC and MSFT. Clearly, the latter are good candidate for inclusion in the portfolio. The exact amount of each stock also depends on the correlation among stocks through the variance of portfolio returns $\sigma^2_p$, and so we need to formalise our objective function under the given constraints.

In order to compute the GA fitness function, we define the following functions:
\begin{example}
> weights <- function(w)      # normalised weights
  { drop(w/sum(w)) }
> ExpReturn <- function(w)    # expected return
  { sum(weights(w)*R) }
> VarPortfolio <- function(w) # objective function
  { 
    w <- weights(w)
    drop(w %*% S %*% w) 
  }
\end{example}

We may define the fitness function to be maximised as the (negative) variance of the portfolio penalised by an amount which is function of the distance between the expected return of the portfolio and the target value:
\begin{example} 
> fitness <- function(w)      # fitness function
  {
    ER <- ExpReturn(w)-0.01
    penalty <- if(ER < 0) 100*ER^2 else 0
    -(VarPortfolio(w) + penalty)
  }
\end{example}
A hybrid GA with local search can be obtained with the following call:
\begin{example}
> GA <- ga(type = "real-valued", fitness = fitness, 
           min = rep(0, nStocks), max = rep(1, nStocks), names = myStocks, 
           maxiter = 1000, run = 200, optim = TRUE)
> summary(GA)
+-----------------------------------+
|         Genetic Algorithm         |
+-----------------------------------+

GA settings: 
Type                  =  real-valued 
Population size       =  50 
Number of generations =  1000 
Elitism               =  2 
Crossover probability =  0.8 
Mutation probability  =  0.1 
Search domain = 
    AAPL XOM GOOGL MSFT GE JNJ WMT CVX PG WFC
Min    0   0     0    0  0   0   0   0  0   0
Max    1   1     1    1  1   1   1   1  1   1

GA results: 
Iterations             = 216 
Fitness function value = -0.00049345 
Solution = 
         AAPL     XOM    GOOGL   MSFT GE JNJ     WMT     CVX      PG    WFC
[1,] 0.031021 0.11981 0.035005 0.5248  0   0 0.17327 0.26192 0.18141 0.9932
> plot(GA)
\end{example}
The last command produces the graph on Figure~\ref{fig2-4:portfolio}c, which shows the trace of best, mean, and median values during the HGA iterations. The vertical dashes at the top of the graph indicate where the local search occurred. 
It is interesting to note that the inclusion of a local search greatly speedup the termination of the GA search, which converges after 216 iterations. Without including the local optimisation step, a fitness function value within a 1\% from the maximum value found above is attained after $1,633$ iterations, whereas the same maximum fitness value cannot be achieved even after $100,000$ iterations.

The estimated portfolio weights and the corresponding expected return and variance are computed as:
\begin{example}
> (w <- weights(GA@solution))
    AAPL      XOM    GOOGL     MSFT       GE      JNJ      WMT      CVX 
0.013369 0.051632 0.015085 0.226166 0.000000 0.000000 0.074671 0.112875 
      PG      WFC 
0.078178 0.428025 
> ExpReturn(w)
[1] 0.016168
> VarPortfolio(w)
[1] 0.00049345
> barplot(w, xlab = "Stocks", ylab = "Portfolio weights", 
          cex.names = 0.7, col = .colorwheelPalette(10))
\end{example} 
The last command draws a barchart of the optimal portfolio selected, and it is shown in Figure~\ref{fig2-4:portfolio}b.

\subsection{Poisson change-point model}

In the study of stochastic processes a common problem is to determine whether or not the functioning of a process has been modified over time. Change-point models assume that such a change is occurring at some point in time in a relatively abrupt manner \citep{Lindsey:2004}.

In a single change-point model the distribution of a response variable $Y_t$ at time $t$ is altered at the unknown point in time $\tau$, so we can write
\begin{equation}
Y_t \sim 
\begin{cases}
f(y_t; \theta_1) & t < \tau \\
f(y_t; \theta_2) & t \ge \tau
\end{cases}
\label{eq:change-point}
\end{equation}
where $f(\cdot)$ is some given parametric distribution depending on $\theta_k$ for $k = \{1,2\}$, and $\tau$ is an unknown parameter giving the change-point time. Some or all of the elements of the vector of parameters $\theta_k$ in model \eqref{eq:change-point} may change over time. In more complex settings, the distribution function itself may be different before and after the change point.

Given a sample $\{y_t; t=1,\ldots,T\}$ of observations over time, the log-likelihood function of the change-point problem is
\begin{equation}
\loglik(\theta_1, \theta_2, \tau; y_1, \ldots, y_T) = 
\sum_{t < \tau} \log f(y_t; \theta_1) +  
\sum_{t \ge \tau} \log f(y_t; \theta_2)
\label{eq:change-point-loglik}
\end{equation} 
Further, for a Poisson change-point model we assume that $f(y_t; \theta_k)$ is the Poisson distribution with mean parameter $\theta_k$.

Consider the British coal-mining disasters dataset which provides the annual counts of disasters (having at least 10 deaths) from 1851 to 1962 \citep{Jarrett:1979, Raftery:Akman:1986}.
The data from Table~1 of \citet{Carlin:Gelfand:Smith:1992} are the following:
\begin{example}
> data <- data.frame(
    y = c(4, 5, 4, 1, 0, 4, 3, 4, 0, 6, 3, 3, 4, 0, 2, 6, 3, 3, 5, 4, 5, 3, 1, 
          4, 4, 1, 5, 5, 3, 4, 2, 5, 2, 2, 3, 4, 2, 1, 3, 2, 2, 1, 1, 1, 1, 3, 
          0, 0, 1, 0, 1, 1, 0, 0, 3, 1, 0, 3, 2, 2, 0, 1, 1, 1, 0, 1, 0, 1, 0, 
          0, 0, 2, 1, 0, 0, 0, 1, 1, 0, 2, 3, 3, 1, 1, 2, 1, 1, 1, 1, 2, 4, 2, 
          0, 0, 0, 1, 4, 0, 0, 0, 1, 0, 0, 0, 0, 0, 1, 0, 0, 1, 0, 1),
    year = 1851:1962,
    t = 1:112)
\end{example}

\begin{figure}[htbp]
\centering\footnotesize
\begin{minipage}{0.49\textwidth}
  \centering
  \includegraphics[width=\textwidth]{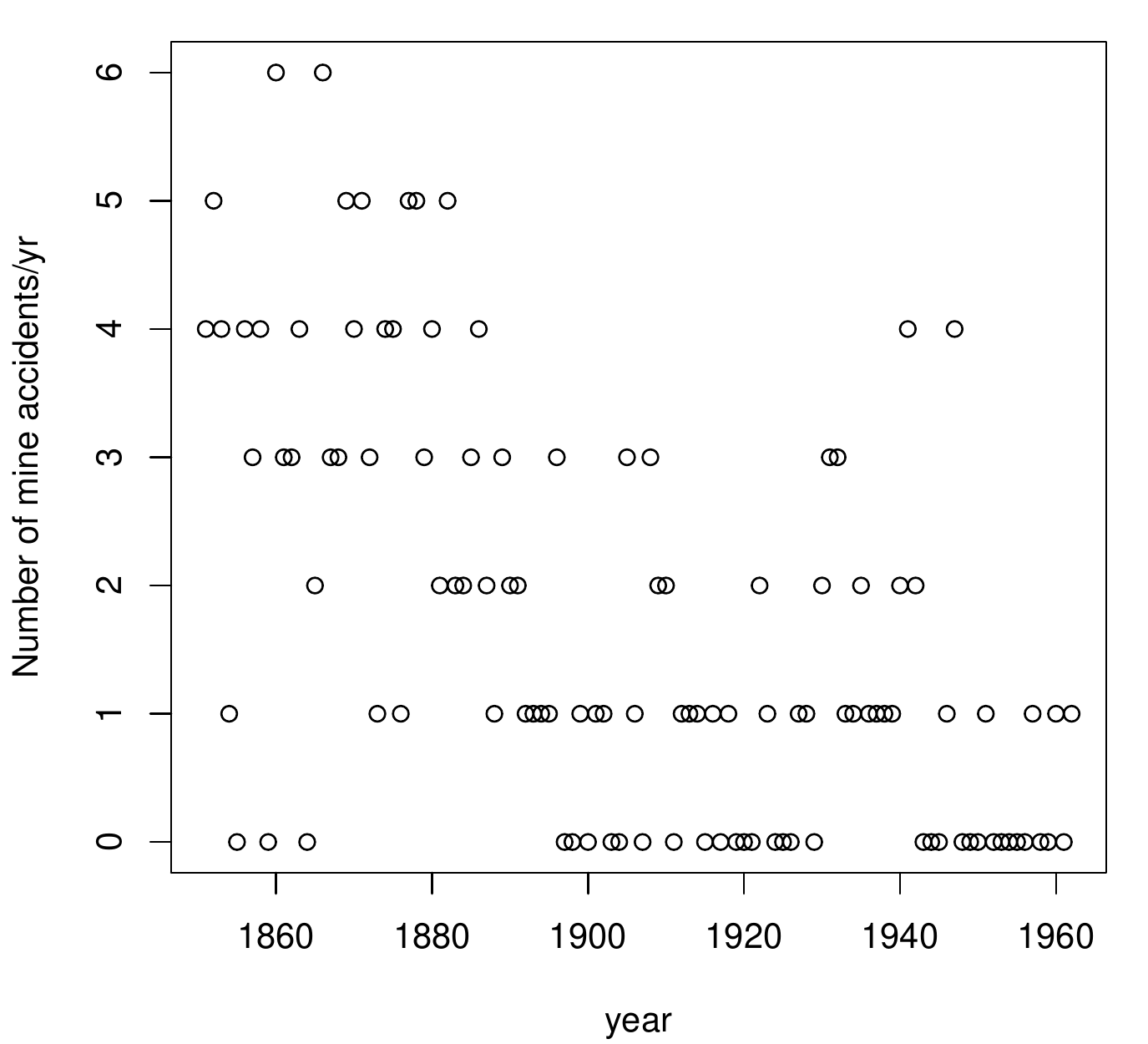}
  (a)
\end{minipage}
\begin{minipage}{0.49\textwidth}
  \centering
  \includegraphics[width=\textwidth]{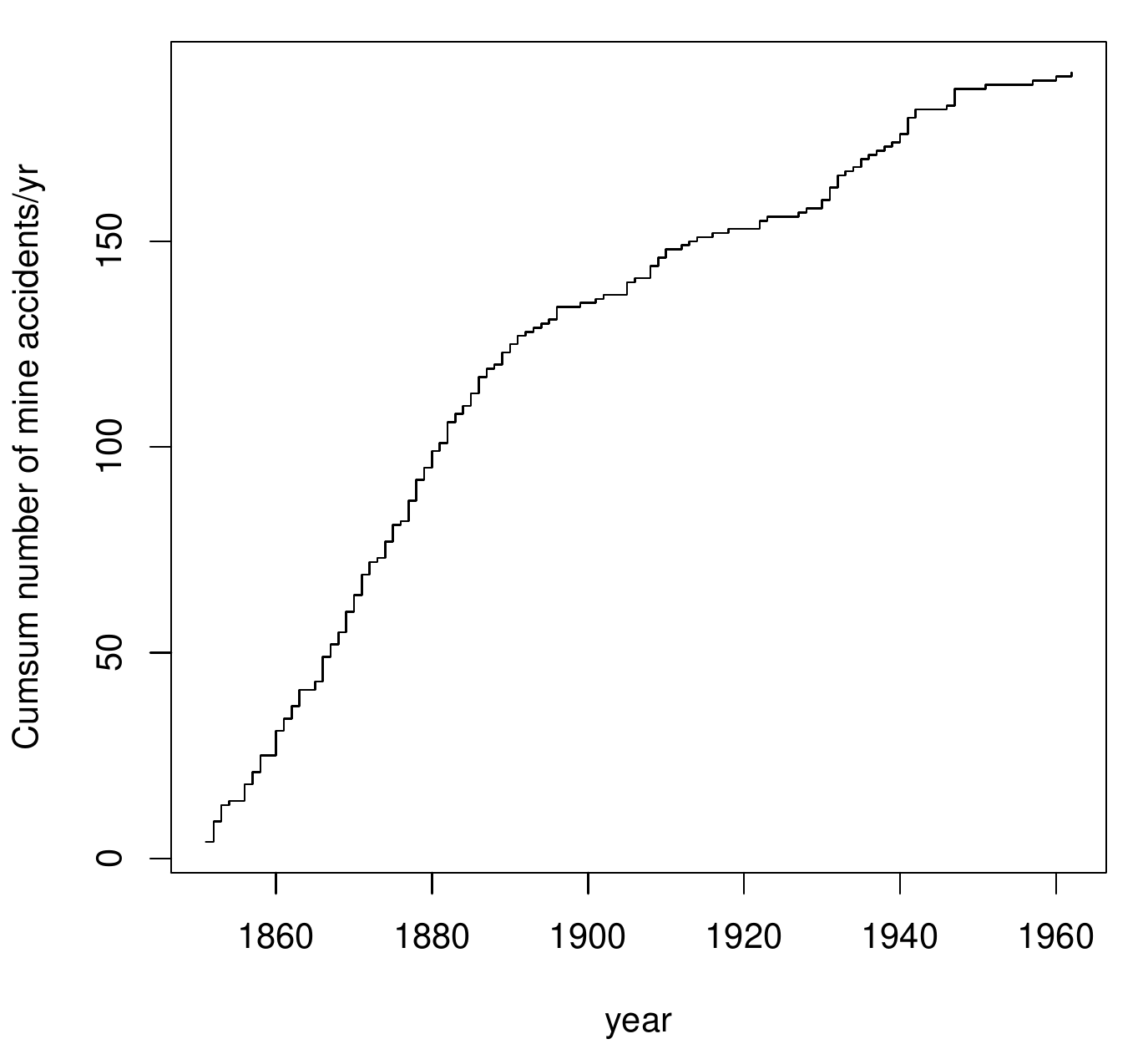}
  (b)
\end{minipage}
  \caption{Plots of the number of yearly coal-mining accidents (a) and cumulative sum of mine accidents (b) from 1851 to 1962 in Great Britain.}
  \label{fig1:coalmine}
\end{figure}

Graphs of annual counts and cumulative sums over time are shown in Figure~\ref{fig1:coalmine}. These can be obtained using the following code:
\begin{example}
> plot(y ~ year, data = data, ylab = "Number of mine accidents/yr")
> plot(cumsum(y) ~ year, data = data, type = "s",
       ylab = "Cumsum number of mine accidents/yr")
\end{example}

Both graphs seems to suggest a two-regime behaviour for the number of coal-mining disasters. 

We start the analysis by fitting a no change-point model, i.e. assuming a homogeneous Poisson process with constant mean. Clearly, in this simple case the MLE of the Poisson parameter is the sample mean of counts. However, for illustrative purposes we write down the log-likelihood and we maximise it with a hybrid GA. 
\begin{example}
> loglik1 <- function(th, data)
 { 
   mu <- exp(th)  # Poisson mean
   sum(dpois(data$y, mu, log = TRUE))
 }
> GA1 <- ga(type = "real-valued", 
            fitness = loglik1, data = data,
            min = log(1e-5), max = log(6), names = "th",
            maxiter = 200, run = 50, 
            optim = TRUE)
> exp(GA1@solution[1,])
1.7054
> mean(data$y)
[1] 1.7054
\end{example}

For the change-point model in \eqref{eq:change-point}, the mean function can be expressed as
$$
\mu_t = \exp\left\{ \theta_1 + (\theta_2 - \theta_1) I(t \ge \tau) \right\},
$$ 
where $\tau$ is the time of change-point, $\theta_1$ is the mean of the first regime, i.e. when $t < \tau$, $\theta_2$ is the mean of the second regime, i.e. when $t \ge \tau$, and $I(\cdot)$ denotes the indicator function (which is equal to 1 if its argument is true and 0 otherwise). In \R\ the above mean function and the log-likelihood from \eqref{eq:change-point-loglik} can be written as
\begin{example}
> meanFun <- function(th, t)
  { 
    tau <- th[3]             # change-point parameter
    th <- th[1:2]            # mean-related parameters
    X <- cbind(1, t >= tau)  # design matrix
    exp(drop(X %*% th))
  }
> loglik2 <- function(th, data)
  { 
    mu <- meanFun(th, data$t)  # vector of Poisson means 
    sum(dpois(data$y, mu, log = TRUE))
  }
\end{example}
The vector \code{th} contains the three parameters that have to be estimated from the sample dataset \code{data}. 
Note that, for convenience, it is defined as $(\theta_1, \theta^*_2, \tau)'$, where $\theta^*_2 = (\theta_2 - \theta_1)$ is the differential mean effect of second regime.

Maximising the log-likelihood in \code{loglik2()} by iterative derivative-based methods is not viable due to lack of differentiability with respect to $\tau$. However, hybrid GAs can be efficiently used in this case as follows:
\begin{example}
> GA2 <- ga(type = "real-valued", 
            fitness = loglik2, data = data,
            min = c(log(1e-5), log(1e-5), min(data$t)), 
            max = c(log(6), log(6), max(data$t)+1),
            names = c("th1", "th2", "tau"),
            maxiter = 1000, run = 200, 
            optim = TRUE)
> summary(GA2)
+-----------------------------------+
|         Genetic Algorithm         |
+-----------------------------------+

GA settings: 
Type                  =  real-valued 
Population size       =  50 
Number of generations =  1000 
Elitism               =  2 
Crossover probability =  0.8 
Mutation probability  =  0.1 
Search domain = 
         th1      th2 tau
Min -11.5129 -11.5129   1
Max   1.7918   1.7918 113

GA results: 
Iterations             = 364 
Fitness function value = -168.86 
Solution = 
        th1     th2    tau
[1,] 1.1306 -1.2344 41.804
> (mean <- exp(cumsum(GA2@solution[1,1:2]))) # mean function parameters
    th1     th2 
3.09756 0.90141 
> (tau <- GA2@solution[1,3])                 # change-point
   tau 
41.804
\end{example}
Note that both the estimated change-point and the means are quite close to those reported by \citet{Raftery:Akman:1986}, and \citet{Carlin:Gelfand:Smith:1992}, using Bayesian methodology.

\begin{figure}[htbp]
\centering\footnotesize
\begin{minipage}{0.49\textwidth}
  \centering
  \includegraphics[width=\textwidth]{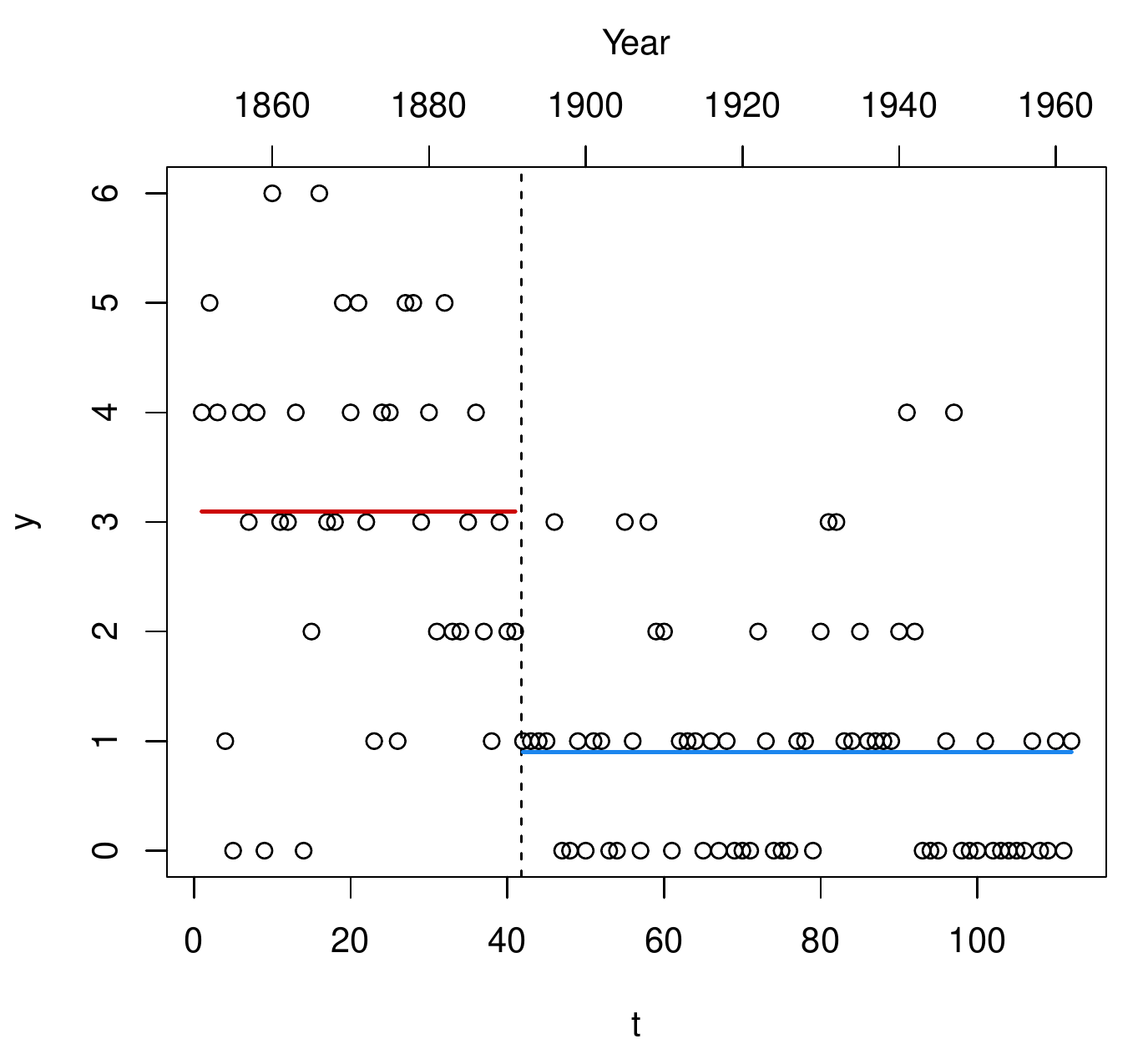}
  (a)
\end{minipage}
\begin{minipage}{0.49\textwidth}
  \centering
  \includegraphics[width=\textwidth]{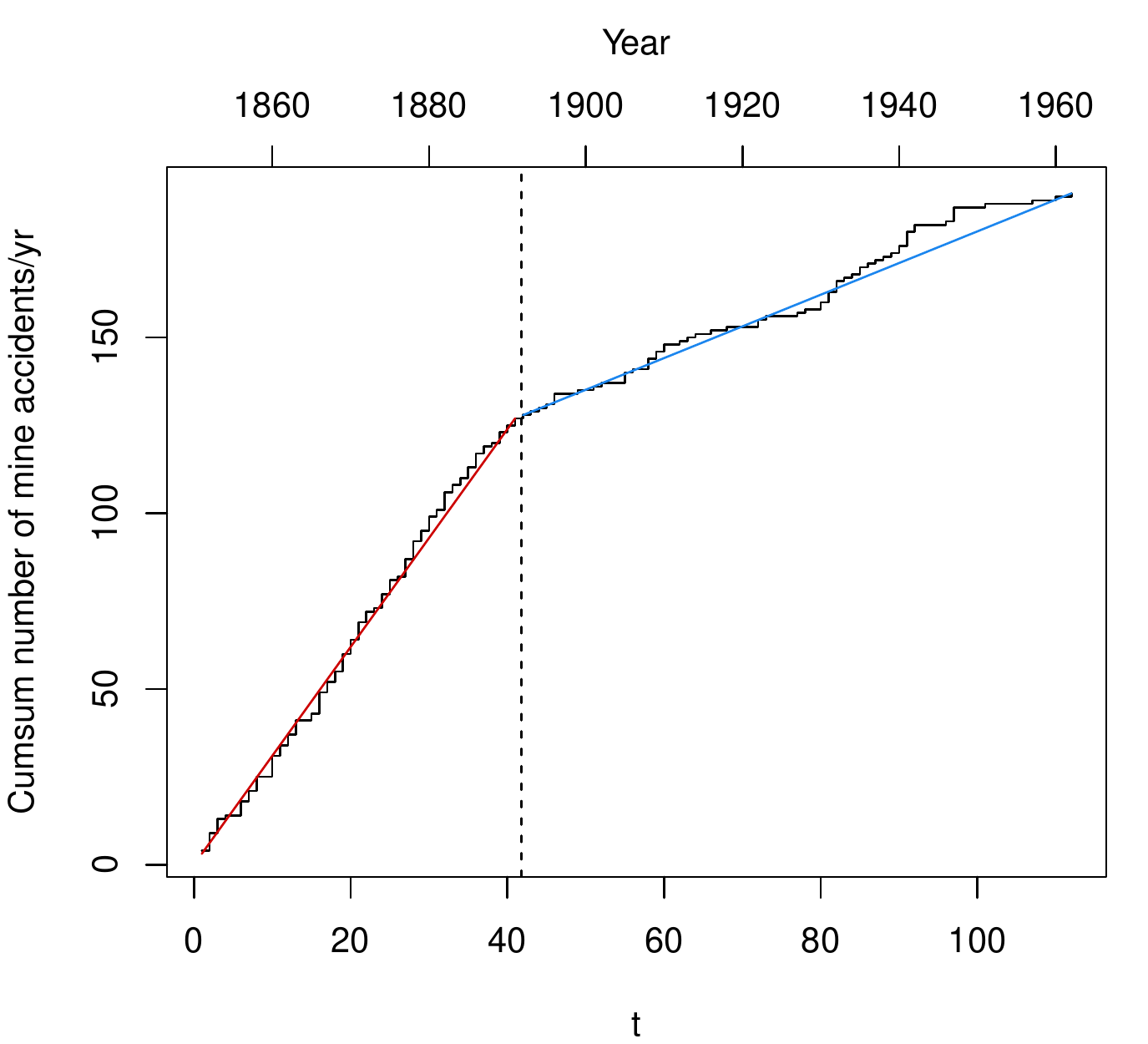}
  (b)
\end{minipage}
  \caption{Summary plots for the change-point model fitted to the British coal-mining accidents dataset: (a) plot of observed counts over time with the estimated means before and after the estimated change-point (vertical dashed line); (b) plot of observed cumulative counts (step function) and the cumulative estimated mean counts.}
  \label{fig2:coalmine}
\end{figure}

The two estimated models can be compared using a model selection criterion, such as the Bayesian information criterion \citep[BIC;][]{Schwartz:1978} defined as 
$$
\BIC = 2 \loglik(\hat{\theta};y) - \nu\log(n)
$$
where $\loglik(\hat{\theta}; y)$ is the log-likelihood evaluated at the MLE $\hat{\theta}$, $n$ is the number of observations, and $\nu$ is the number of estimated parameters. Using this definition, larger values of BIC are preferable.
\begin{example}
> (tab <- data.frame(
    loglik = c(GA1@fitnessValue, GA2@fitnessValue),
    df = c(ncol(GA1@solution), ncol(GA2@solution)),
    BIC = c(2*GA1@fitnessValue - log(nrow(data))*ncol(GA1@solution),
            2*GA2@fitnessValue - log(nrow(data))*ncol(GA2@solution))))

   loglik df     BIC
1 -203.86  1 -412.43
2 -168.86  3 -351.88
\end{example}
A comparison of BIC values clearly indicates a preference for the change-point model.
We may summarise the estimated model by drawing a graph of observed counts over time with the estimated means before and after the change-point:
\begin{example}
> mu <- meanFun(GA2@solution, data$t)
> col <- c("red3", "dodgerblue2")
> with(data, 
  { plot(t, y)
    abline(v = tau, lty = 2)
    lines(t[t < tau], mu[t < tau], col = col[1], lwd = 2)
    lines(t[t >= tau], mu[t >= tau], col = col[2], lwd = 2)
    par(new=TRUE)
    plot(year, cumsum(y), type = "n", axes = FALSE, xlab = NA, ylab = NA)
    axis(side = 3); mtext("Year", side = 3, line = 2.5)
  })
\end{example} 
and a graph of observed cumulative counts and the estimated cumulative mean counts:
\begin{example}
> with(data, 
  { plot(t, cumsum(y), type = "s", ylab = "Cumsum number of mine accidents/yr")
    abline(v = tau, lty = 2)
    lines(t[t < tau], cumsum(mu)[t < tau], col = col[1], lwd = 2)
    lines(t[t >= tau], cumsum(mu)[t >= tau], col = col[2], lwd = 2)
    par(new=TRUE)
    plot(year, cumsum(y), type = "n", axes = FALSE, xlab = NA, ylab = NA)
    axis(side = 3); mtext("Year", side = 3, line = 2.5)
  })
\end{example}
Both graphs are reported in Figure~\ref{fig2:coalmine}. The latter plot is particularly illuminating of the good fit achieved by the selected model.

\subsection{S-I-R model for influenza epidemic}

The S-I-R model is a simple epidemiology compartmental model proposed by \citet{Kermack:McKendrick:1927}, which assumes a fixed population with only three compartments or states:
\begin{itemize}
\item $S(t)$ = number of susceptible, i.e. the number of individuals susceptible to the disease not yet infected at time $t$;
\item $I(t)$ = number of infected, i.e. the number of individuals who have been infected at time $t$ with the disease and are capable of spreading the disease to those in the susceptible category;
\item $R(t)$ = number of  recovered, i.e. those individuals who have been infected and then removed from the disease, either due to immunisation or due to death. Members of this compartment are not able to be infected again or to transmit the infection to others.
\end{itemize}

Using a fixed population, i.e. with constant size $N = S(t) + I(t) + R(t)$, \citet{Kermack:McKendrick:1927} derived the following system of quadratic ODEs:
\begin{align*}
\frac{dS}{dt} & = - \beta S I \\
\frac{dI}{dt} & = \beta S I - \gamma I \\
\frac{dR}{dt} & = \gamma I
\end{align*}
where $\beta > 0$ is the rate (constant for all individuals) at which an infected person infects a susceptible person, and $\gamma > 0$ is the rate at which infected people recover from the disease. 

The flow of the S-I-R model can be represented in the following scheme:
$$
\fbox{$S(t)$} \quad\xrightarrow{\beta S I}\quad
\fbox{$I(t)$} \quad\xrightarrow{\gamma I}\quad
\fbox{$R(t)$}
$$
where boxes represent the compartments and arrows indicate flows between compartments. Note that 
$\frac{dS}{dt} + \frac{dI}{dt} + \frac{dR}{dt} = 0$, then $S(t) + I(t) + R(t) = N$, and the initial condition $S(0) > 0, I(0) > 0, R(0) = 0$.
Thus, the system can be reduced to a system of two ODEs. 

For our data analysis example, we consider the influenza epidemic in an English boarding school from 22nd January to 4th February 1978 as described in \citet[p. 325--326]{Murray:2002}. 
There were 763 resident boys in the school, and one (the initial infective) returned from winter break with illness. Over the course of 13 days, 512 boys were infected by the flu.
\begin{example}
> day <- 0:14
> Infected <- c(1,3,6,25,73,222,294,258,237,191,125,69,27,11,4)
> N <- 763
> init <- c(S = N-1, I = 1, R = 0)
> plot(day, Infected)
\end{example}

We aim at estimating the values of $\beta$ and $\gamma$ based on the observed data by minimising the following loss function:
\begin{equation}
RSS(\beta, \gamma) = \sum_t e(t)^2 = \sum_t \left( I(t) - \hat{I}(t) \right)^2,
\label{eq:RSS_sir}
\end{equation}
where $I(t)$ is the number of infected observed at time $t$, and $\hat{I}(t)$ is the corresponding number of infected predicted by the model, which depends on the unknown parameters $\beta$ and $\gamma$. Nonlinear least squares can be used to fit this model to data, but it strongly depends on the initial values as shown below. A more robust approach can be pursued by using GAs.

First of all, we need to define a function which computes the values of the derivatives in the ODE system at time $t$. This function is then used, together with the initial values of the system and the time sequence, by function \code{ode()} in the \R\ package \CRANpkg{deSolve} to solve the ODE system:
\begin{example}
> library(deSolve)
> SIR <- function(time, state, parameters) 
  {
    par <- as.list(c(state, parameters))
    with(par, { dS <- -beta * S * I
                dI <- beta * S * I - gamma * I
                dR <- gamma * I
                list(c(dS, dI, dR)) 
              })
  }
> RSS.SIR <- function(parameters)
  {
    names(parameters) <- c("beta", "gamma")
    out <- ode(y = init, times = day, func = SIR, parms = parameters)
    fit <- out[,3]
    RSS <- sum((Infected - fit)^2)
    return(RSS)
  }
\end{example}
The function \code{RSS.SIR()} computes the predicted number of infected $\hat{I}(t)$ from the solution of ODE system for the input \code{parameters} values, and returns the objective function in \eqref{eq:RSS_sir} to be minimised. 
Then, a \code{ga()} function call can be used with local search to find the optimal values of parameters $(\beta, \gamma)$ in S-I-R model. Note that the fitness function is specified as a local function which simply returns the negative of the objective function. 
In this case, fine tuning of local search is specified through the optional argument \code{optimArgs}: the selection pressure is set with \code{pressel} at a higher value, so better solutions have higher probability of being used as starting point for the local search, and \code{maxit} gets a two-values vector specifying the maximum number of iterations to be used, respectively, during the GA evolution and after the final iteration.
\begin{example}
> GA <- ga(type = "real-valued", 
           fitness = function(par) -RSS.SIR(par),
           min = c(0,0), max = c(0.1,0.5), 
           names = c("beta", "gamma"),
           popSize = 25, maxiter = 1000, run = 200,
           optim = TRUE, 
           optimArgs = list(pressel = 0.8, 
                            control = list(maxit = c(10,100))))
> summary(GA)
+-----------------------------------+
|         Genetic Algorithm         |
+-----------------------------------+

GA settings: 
Type                  =  real-valued 
Population size       =  25 
Number of generations =  1000 
Elitism               =  1 
Crossover probability =  0.8 
Mutation probability  =  0.1 
Search domain = 
    beta gamma
Min  0.0   0.0
Max  0.1   0.5

GA results: 
Iterations             = 503 
Fitness function value = -4507.1 
Solution = 
          beta   gamma
[1,] 0.0021806 0.44516
\end{example}

Based on the estimated parameters other quantities of interest can be computed. For instance, $1/\gamma = 1/0.44516 \approx 2.25$ is the average recovery time which expresses the duration of infection (in days), and $\beta/\gamma \times 100 = 0.0021806/0.44516 \times 100 \approx 0.49\%$ is the infection's contact rate.

\begin{figure}[htbp]
\centering\footnotesize
\begin{minipage}{0.49\textwidth}
  \centering
  \includegraphics[width=\textwidth]{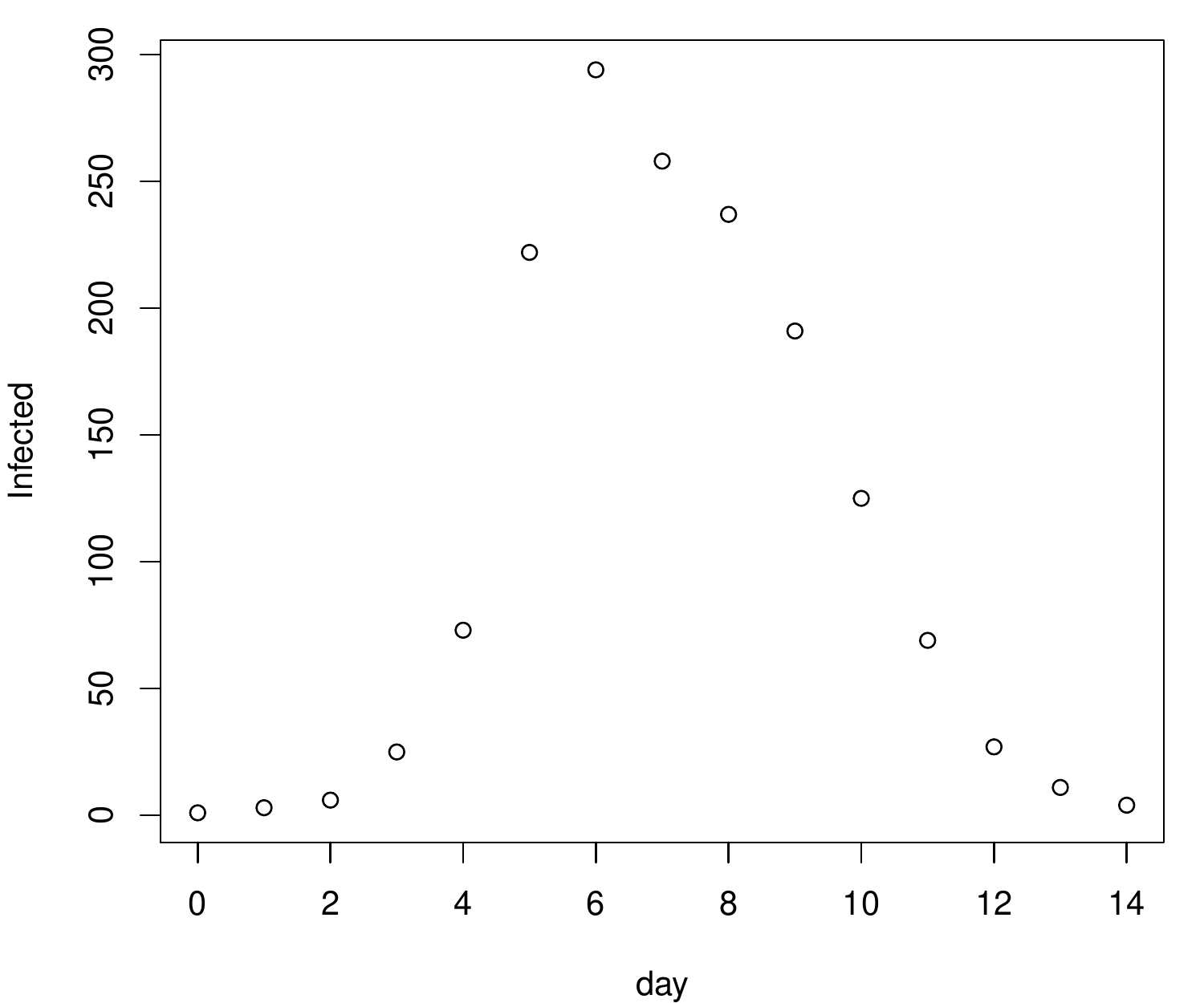}
  (a)
\end{minipage}
\begin{minipage}{0.49\textwidth}
  \centering
  \includegraphics[width=\textwidth]{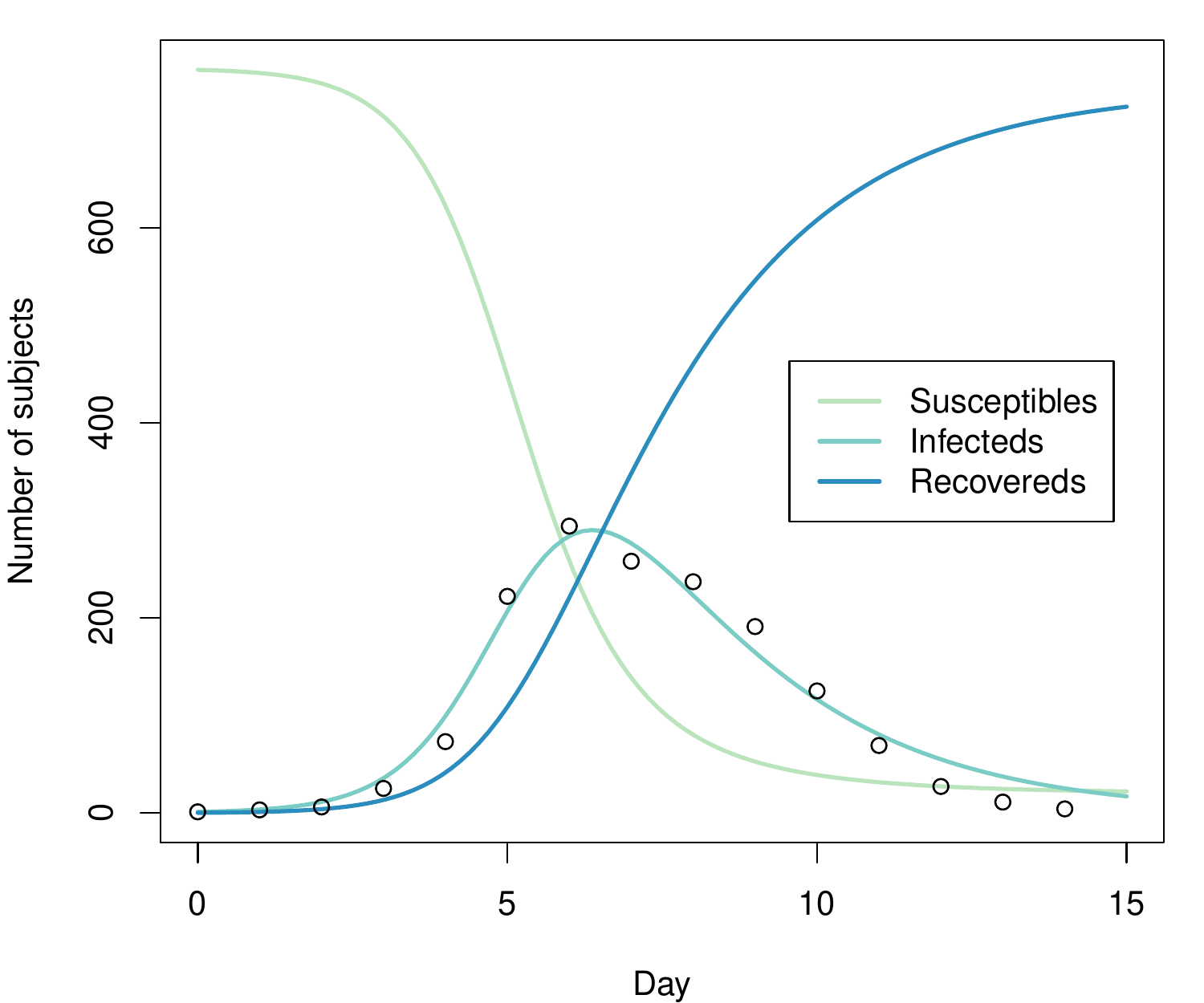}
  (b)
\end{minipage}
  \caption{Influenza epidemic in an English boarding school in winter 1978: (a) plot of the number of infected students; (b) model predictions from the S-I-R model with parameters estimated by hybrid GAs.}
  \label{fig1:sir}
\end{figure}

The graph in Figure~\ref{fig1:sir}b provides a graphical summary of quantities involved in S-I-R model and the dynamic evolution of epidemia:
\begin{example}
> t <- seq(0, 15, length = 100)
> fit <- data.frame(ode(y = init, times = t, func = SIR, 
                        parms = GA@solution[1,]))
> col <- brewer.pal(4, "GnBu")[-1]
> matplot(fit$time, fit[,2:4], type = "l", 
          xlab = "Day", ylab = "Number of subjects",
          lwd = 2, lty = 1, col = col)
> points(day, Infected)
> legend("right", c("Susceptibles", "Infecteds", "Recovereds"), 
         lty = 1, lwd = 2, col = col, inset = 0.05)
\end{example}

We note that \citet{Murray:2002} reported solution $(\beta = 0.00218, \gamma = 0.441)$ gives a RSS equal to $4535.9$, larger than the optimal solution found by HGAs which is equal to $4507.1$. Furthermore, direct optimisation depends on starting values and often converges to sub-optimal solutions as, for instance, the following:
\begin{example}
> optim(c(0.001,0.4), RSS.SIR, method = "L-BFGS-B", lower = GA@min, upper = GA@max)
$par
[1] 0.0021434 0.3954033

$value
[1] 8764.9

$counts
function gradient 
      96       96 

$convergence
[1] 52

$message
[1] "ERROR: ABNORMAL_TERMINATION_IN_LNSRCH"
\end{example}

\section{Parallel genetic algorithms}
\label{sec:PGA}

Parallel computing in its essence involves the simultaneous use of multiple computing resources to solve a computational problem.
This is viable when a task can be divided into several parts that can be solved simultaneously and independently, either on a single multi-core processors machine or on a cluster of multiple computers.

Support for parallel computing in \R\ is available since 2011 (version 2.14.0) through the base package \pkg{parallel}. 
This provides parallel facilities previously contained in packages \pkg{multicore} and \pkg{snow}. Several approaches to parallel computing are available in \R\ \citep{McCallum:Weston:2011}, and an extensive and updated list of \R\ packages is reported in the CRAN task view on \textit{High-Performance and Parallel Computing with \R} \citep{CRAN:HighPerfParComp}.

GAs are regarded as ``embarrassingly parallel'' problems, meaning that they require a large number of independent calculations with negligible synchronisation and communication costs. 
Thus, GAs are particularly suitable for parallel computing, and it is not surprising that such idea has been often exploited to speed up computations (see for instance \citet{Whitley:1994} in the statistical literature). 

\begin{figure}[htb]
\centering
\begin{minipage}[b]{.45\linewidth}
  \centering
  \includegraphics[width=\textwidth]{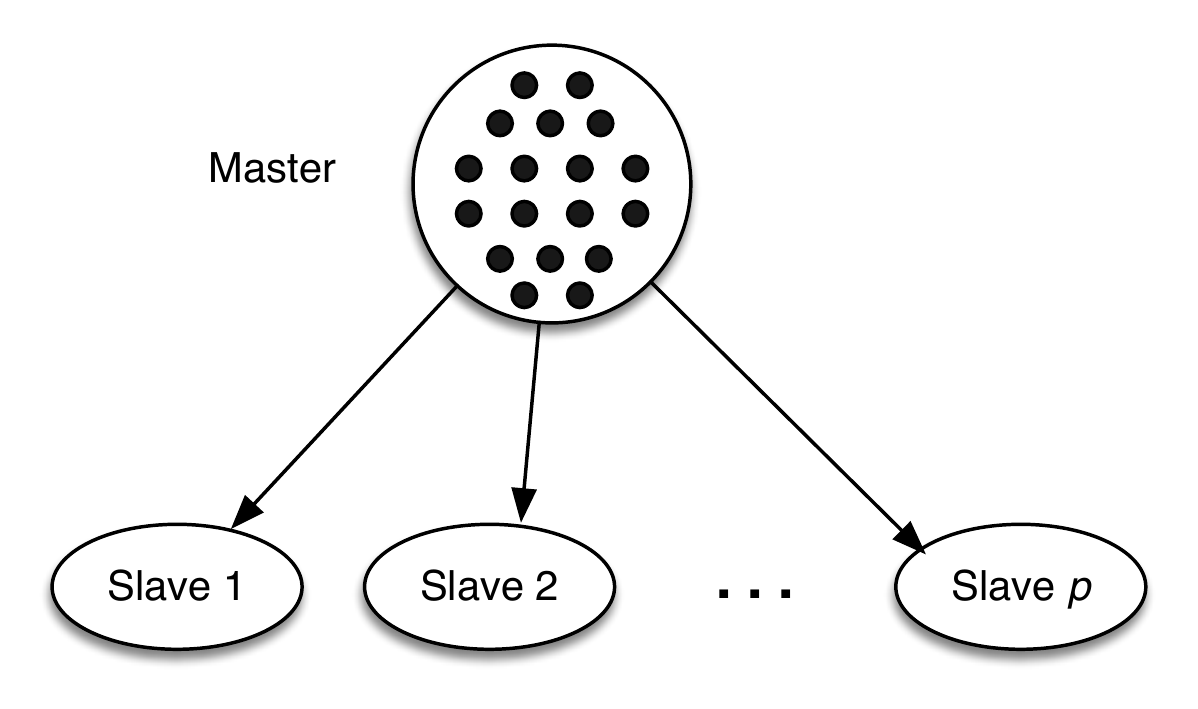}
  \vspace{2ex}
\end{minipage}%
\hspace{1cm}%
\begin{minipage}[b]{.45\linewidth}
  \centering
  \includegraphics[width=\textwidth]{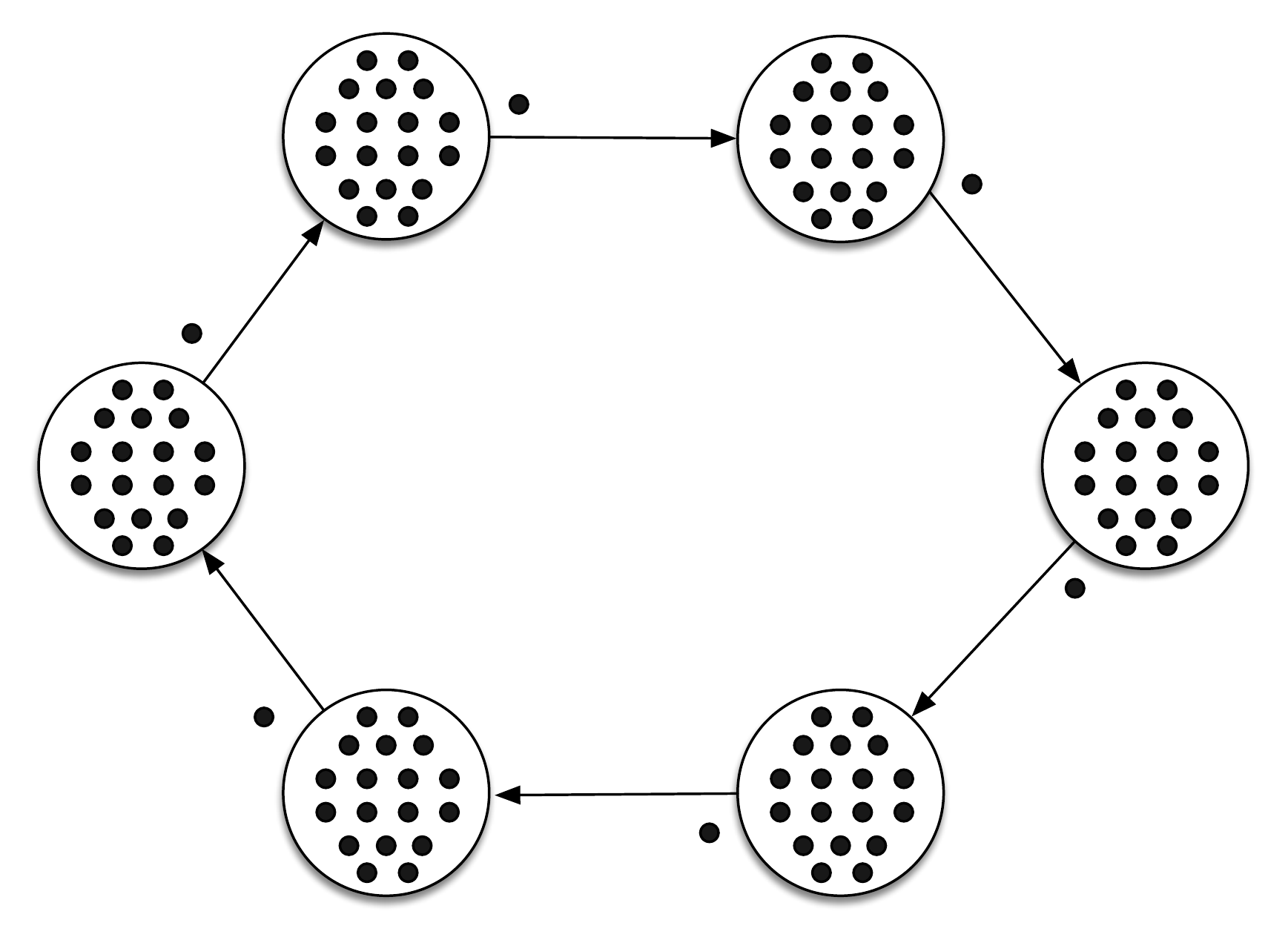}
\end{minipage}\\[-10pt]
\begin{minipage}[t]{.45\linewidth}
  \caption{Master-slaves or global parallel GA scheme (GPGA). The master process stores the population, executes genetic operations, and distributes individuals to the slaves, which only evaluate the fitness of individuals.} 
  \label{fig:GPGA}
\end{minipage}%
\hspace{1cm}%
\begin{minipage}[t]{.45\linewidth}
  \caption{Islands parallel GA scheme (ISLPGA). In a multiple-population parallel GA each process is a simple GA which evolves independently. Individuals occasionally migrate between one island and its neighbours.} 
  \label{fig:ISLPGA}
\end{minipage}%
\end{figure}

\citet{Luque:2011} identify several types of parallel GAs. In the master-slaves approach there is a single population, as in sequential GAs, but the evaluation of fitness is distributed among several processors (\textit{slaves}). The \textit{master} process is responsible of the distribution of the fitness function evaluation tasks performed by the slaves, and for applying genetic operators such as selection, crossover, and mutation (see Figure~\ref{fig:GPGA}). Since the latter operations involve the entire population, it is also known as global parallel GAs (GPGA). This approach is generally efficient when the computational time involving the evaluation of the fitness function is more expensive than the communication overhead between processors. 
 
Another approach is the case of distributed multiple-population GAs, where the population is partitioned into several subpopulations and assigned to separated islands. Independent GAs are executed in each island, and only occasionally sparse exchanges of individuals are performed among these islands (see Figure~\ref{fig:ISLPGA}). This process, called migration, introduces some diversity into the subpopulations, thus preventing the search from getting stuck in local optima. In principle islands can evolve sequentially, but increased computational efficiency is obtained by running GAs in each island in parallel. This approach is known as coarse-grained GAs or island parallel GAs (ISLPGA).

By default, searches performed with the \pkg{GA} package occur sequentially. In some cases, particularly when the evaluation of the fitness function is time consuming, parallelisation of the search algorithm may be able to speed up computing time. 
Starting with version 2.0, the \pkg{GA} package provides facilities for using parallel computing in genetic algorithms following the GPGA approach. 
Recently, with version 3.0, the ISLPGA model has also been implemented in the \pkg{GA} package. The following subsections describes usage of both approaches.

Parallel computing in  the \pkg{GA} package requires the following packages to be installed: \pkg{parallel} (available in base \R), \CRANpkg{doParallel}, \CRANpkg{foreach}, and \CRANpkg{iterators}. Moreover, \CRANpkg{doRNG} is needed for reproducibility in the ISLPGA case.

\subsection{Global parallel implementation}

The GPGA approach to parallel computing in \pkg{GA} can be easily obtained by manipulating the optional argument \code{parallel} in the \code{ga()} function call. This argument accepts several different values. A logical value may be used to specify if parallel computing should be used (\code{TRUE}) or not (\code{FALSE}, default) for evaluating the fitness function. A numeric value can also be supplied, in which case it gives the number of cores/processors to employ; by default, all the available cores, as provided by \code{detectCores()}, are used. 

Two types of parallel functionalities are available depending on system OS: on Windows only \textit{snow} type functionality is present, whereas on POSIX operating systems, such as Unix, GNU/Linux, and Mac OSX, both \textit{snow} and \textit{multicore} (default) functionalities are available. In the latter case, a string can be used as the argument to \code{parallel} to set out which parallelisation tool should be used.

A final option is available if a researcher plans to use a cluster of multiple machines. In this case, \code{ga()} can be executed in parallel using all, or a subset of, the cores available to each machine assigned to the cluster. However, this option requires more work from the user, who needs to set up and register a parallel back end. The resulting cluster object can be passed as input value to the \code{parallel} argument.

\subsection{Islands parallel implementation}

The ISLPGA approach to parallel computing in \pkg{GA} has been implemented in the \code{gaisl()} function. This function accepts the same input arguments as the \code{ga()} function \citep[see][Section~3]{Scrucca:2013}, with the following additional arguments:

\vspace{\dimexpr-2\parsep-2\parskip\relax}%
\begin{center}
\begin{tabular}[t]{lp{0.75\textwidth}}
\code{numIslands} & 
An integer value which specifies the number of islands to use in the genetic evolution (by default is set to $4$). \\
\code{migrationRate} & 
A value in the range $(0,1)$ which gives the proportion of individuals that undergo migration between islands in every exchange (by default equal to $0.10$). \\
\code{migrationInterval} & 
An integer value specifying the number of iterations at which exchange of individuals takes place. This interval between migrations is called an \emph{epoch}, and it is set at 10 by default.
\end{tabular}
\end{center}

The implemented ISLPGA uses a simple \textit{ring topology}, in which each island is connected unidirectionally with another island, hence forming a single continuous pathway (see Figure~\ref{fig:ISLPGA}). Thus, at each exchange step the top individuals, selected according to the specified \code{migrationRate}, substitute random individuals (with the exception of the elitist ones) in the connected island. 

By default, the function \code{gaisl()} uses \code{parallel = TRUE}, i.e. the islands algorithm is run in parallel, but other values can also be provided as described in the previous subsection. Note that it is possible to specify a number of islands larger than the number of available cores. In such a case, the parallel algorithm will be run using blocks of islands, with the block size depending on the maximal number of cores available or the number of processors as specified by the user. 

It has been noted that using parallel islands GAs often leads to, not only faster algorithms, but also superior numerical performance even when the algorithms run on a single processor. This because each island can search in very different regions of the whole search space, thus enhancing the exploratory attitude of evolutionary algorithms. 

\subsection{Simulation study}

In this Section results from a simulation study are presented and discussed. The main goal is to compare the performance of sequential GAs with the two forms of parallel algorithms implemented in the \pkg{GA} package, namely GPGA and ISLPGA, for varying number of cores and different fitness computing times.
A fictitious fitness function is used to allow for controlling the computing time required at each evaluation. This is achieved by including the argument \code{pause} which suspend the execution for a specified time interval (in seconds):
\begin{example}
> fitness <- function(x, pause = 0.1)
  {
    Sys.sleep(pause)
    x*runif(1)
  }
\end{example}
The simulation design parameters used are the following:
\begin{example}
> ncores <- c(1, 2, 4, 8, 16)    # number of cores/processors
> pause  <- c(0.01, 0.1, 1, 2)   # pause during fitness evaluation
> nrep   <- 10                   # number of simulation replications
\end{example}
Thus, \code{ncores} specifies that up to 16 cores or CPU processors are used in the parallel GAs solutions for increasing time spent on fitness evaluation as specified by \code{pause} (in seconds). Each combination of design parameters is replicated \code{nrep = 10} times and results are then averaged. 

GAs are run under the GPGA approach using \code{popSize = 50} and {maxiter = 100}. For ISLPGA runs the \code{numIslands} argument is set at the specified number of cores, with \code{popSize = 160} and \code{maxiter = 100}. The increased population size allows to work with at least 10 individuals on each island when \code{numIslands} is set at the maximum number of cores. In both cases, the remaining arguments in \code{ga()} or \code{gaisl()} function are set at their defaults. 

The study was performed on a 16 cores Intel\textsuperscript{\textregistered} Xeon\textsuperscript{\textregistered} CPU E5-2630 running at 2.40GHz and with 128GB of RAM.
The \R\ code used in the simulation study is provided in the accompanying supplemental material. 

\begin{figure}[!htbp]
\centering
\includegraphics[width=\textwidth]{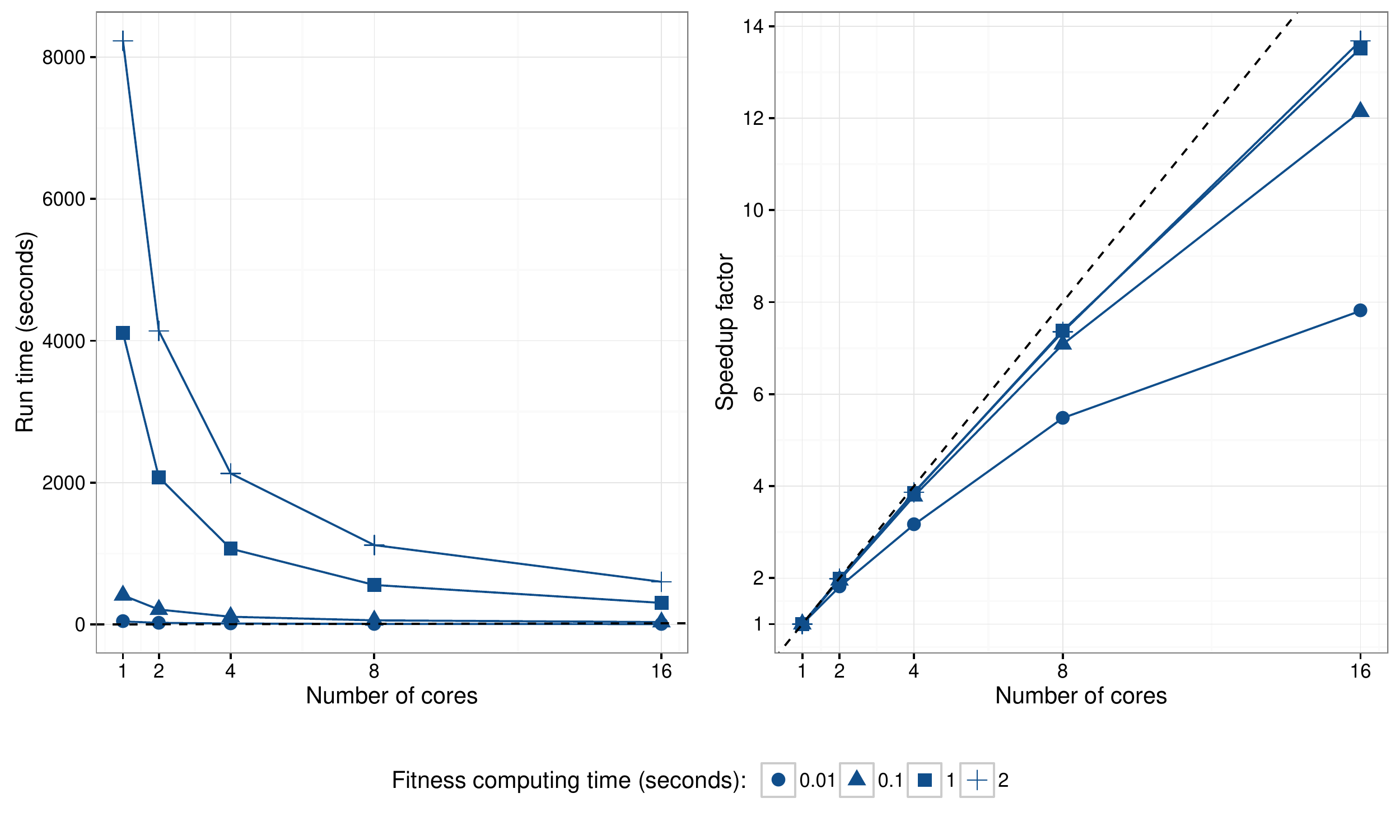}
\caption{Empirical GPGA performance for varying number of cores/processors and different fitness computing times. Graph on the left panel shows the average running times, whereas graph on the right panel shows the speedup factor compared to the sequential run (i.e. when only 1 core is used). In the latter plot, the dashed line represents the ``ideal'' linear speedup.}
\label{fig1:PGA}
\end{figure}

\begin{figure}[!htb]
\centering
\includegraphics[width=\textwidth]{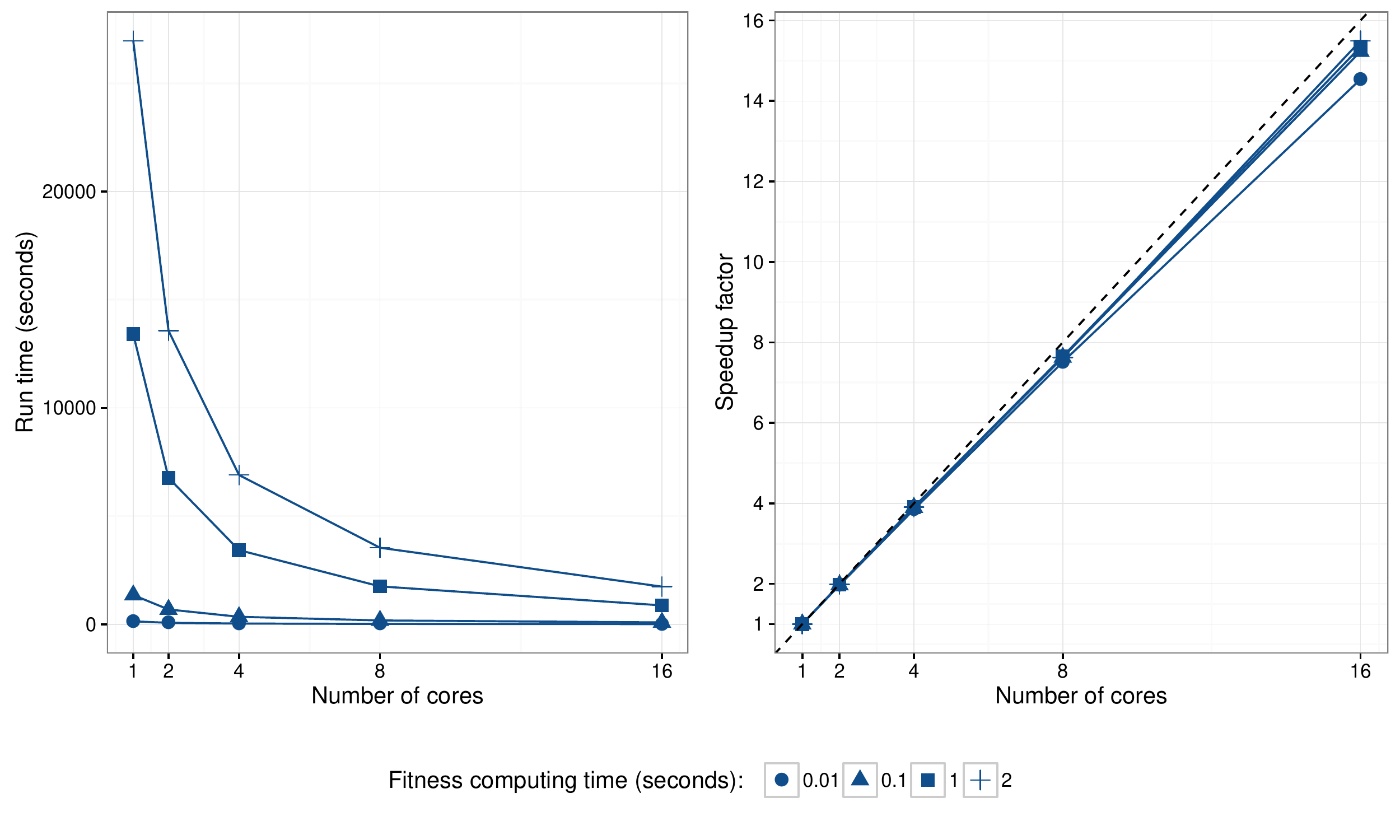}
\caption{Empirical ISLPGA performance for varying number of cores/processors and different fitness computing times. Graph on the left panel shows the average running times, whereas graph on the right panel shows the speedup factor compared to the sequential run (i.e. when only 1 core is used). In the latter plot, the dashed line represents the ``ideal'' linear speedup.} 
\label{fig1:ISLPGA}
\end{figure}

Graphs in the left panel of Figures \ref{fig1:PGA} and \ref{fig1:ISLPGA} show the average execution times needed for varying number of cores and different fitness computing times. As expected, increasing the number of cores allows to run GAs faster, but the improvement is not linear, in particular for the GPGA approach.

By using a machine with $P$ cores/processors, we would like to obtain an increase in calculation speed of $P$ times. However, this is typically not the case because in the implementation of a parallel algorithm there are some inherent non-parallelisable parts and communication costs between tasks \citep{Nakano:2012}.
The speedup achieved using $P$ processors is computed as $s_P = t_1/t_P$, where $t_i$ is the execution time spent using $i$ cores. 
Graphs in the right panel of Figures \ref{fig1:PGA} and \ref{fig1:ISLPGA} show the speedup obtained in our simulation study. For the GPGA approach the speedup is quite good but it is always sub-linear, in particular for the less demanding fitness evaluation time and when the number of cores increases. On the other hand, the ISLPGA implementation shows a very good speedup (nearly linear).  

\begin{figure}[!htb]
\centering
\includegraphics[width=\textwidth]{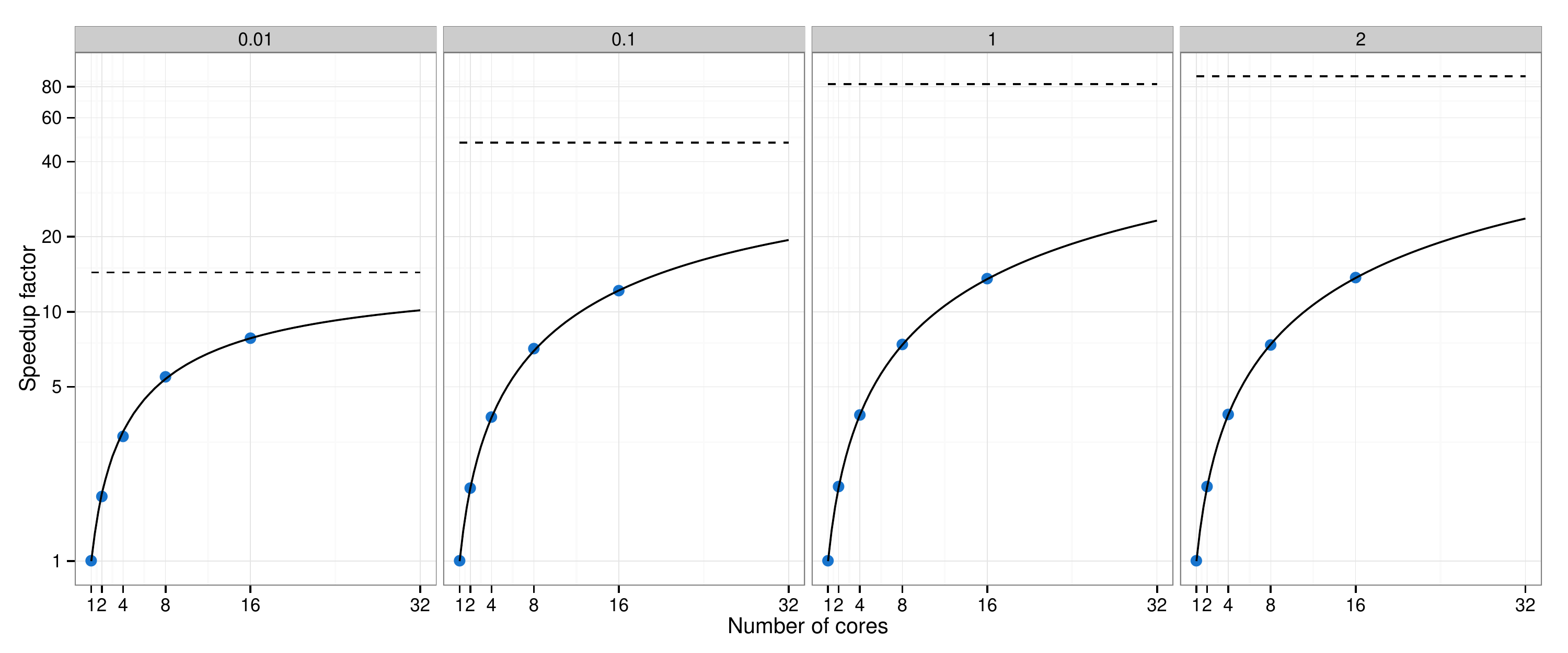}
\caption{Amdahl's law curves for the GPGA approach. Points refer to the speedup factors observed using different number of cores/processors, whereas the curves are estimated using nonlinear least squares. Horizontal dashed lines refer to the maximum speedup theoretically attainable. Each panel corresponds to a different fitness computing time (in seconds), and vertical axes are on log scale.}
\label{fig2:PGA}
\end{figure}

\begin{figure}[!htb]
\centering
\includegraphics[width=\textwidth]{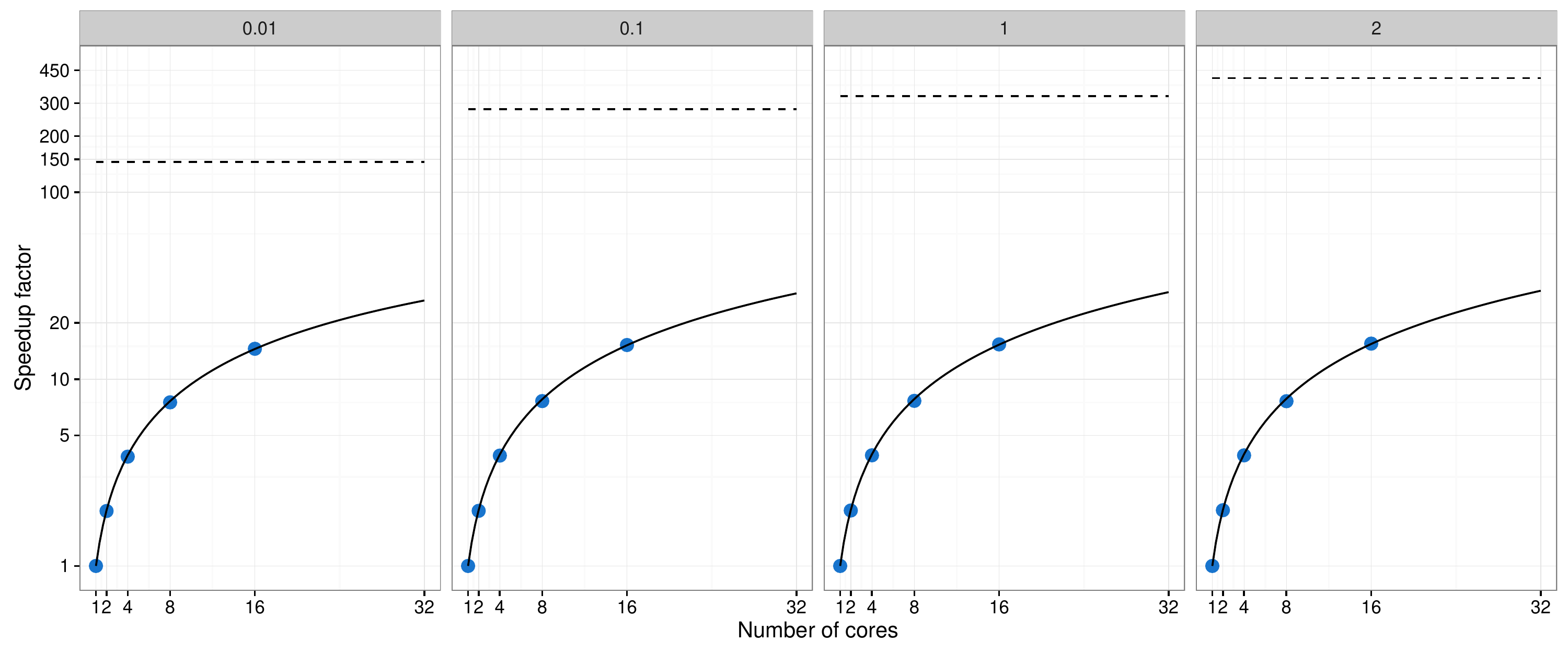}
\caption{Amdahl's law curves for the ISLPGA approach. Points refer to the speedup factors observed using different number of cores/processors, whereas the curves are estimated using nonlinear least squares. Horizontal dashed lines refer to the maximum speedup theoretically attainable. Each panel corresponds to a different fitness computing time (in seconds), and vertical axes are on log scale.}
\label{fig2:ISLPGA}
\end{figure}

Amdahl's law \citep{Amdahl:1967} is often used in parallel computing to predict the theoretical maximum speedup when using multiple processors. According to this, if $f$ is the fraction of non-parallelisable task, i.e. the part of the algorithm that is strictly serial, and $P$ is the number of processors in use, then the speedup obtained on a parallel computing platform follows the equation
\begin{equation}
S_P = \frac{1}{f + (1-f)/P} .
\label{eq:AmdahlLaw}
\end{equation}
In the limit, the above ratio converges to $S_{\max} = 1/f$, which represents the maximum speedup attainable in theory, i.e. by a machine with an infinite number of processors.
Figures \ref{fig2:PGA} and \ref{fig2:ISLPGA} show the observed speedup factors $S_P$ and the estimated Amdahl's law curves fitted by nonlinear least squares. In all the cases, Amdahl's law appears to well approximate the observed behaviour. The horizontal dashed lines are drawn at the maximum speedup $S_{\max}$, which is computed based on the estimated fraction of non-parallelisable task $f$ (see also Table~\ref{tab1:ISLPGA}). As the time required for evaluating the fitness function increases, the maximum speedup attainable also increases. As noted earlier, the ISLPGA approach shows an improved efficiency compared to the simple GPGA. 

\begin{table}[!htb]
\centering
\caption{Fraction of non-parallelisable task $(f)$ estimated by nonlinear least squares using the Amdahl's law, and corresponding theoretical speedup $(S_{\max})$ for the GPGA and ISLPGA approaches.}
\label{tab1:ISLPGA}
\begin{tabular}{lrrrrcrrrr}
\toprule
 & \multicolumn{4}{c}{GPGA} && \multicolumn{4}{c}{ISLPGA} \\
 \cline{2-5} \cline{7-10} \\[-2ex]
 & 0.01 & 0.1 & 1 & 2 && 0.01 & 0.1 & 1 & 2\\
\cline{2-5} \cline{7-10} \\[-2ex]
$f$        & 0.0695 & 0.0209 & 0.0122 & 0.0114 & 
           & 0.0069 & 0.0036 & 0.0031 &  0.0025 \\
$S_{\max}$ & 14.38  &  47.76 &  81.88 &  87.88 & 
           & 145.29 & 278.57 & 327.12 & 408.58 \\
\bottomrule
\end{tabular}
\end{table}

\subsection{ARIMA order selection}

Autoregressive moving average (ARMA) models are a broad class of parametric models for stationary time series popularised by \citet{Box:Jenkins:1976}. They provide a parsimonious description of a stationary stochastic process in terms of two polynomials, one for the auto-regression and the second for the moving average. 
Nonstationay time series can be modelled by including an initial differencing step (``integrated'' part of the model). This leads to autoregressive integrated moving average (ARIMA) models, a popular modelling approach in real-world processes.

ARIMA models can be fitted by MLE after identifying the order $(p,d,q)$ for the autoregressive, integrated, and moving average components, respectively. This is typically achieved by preliminary inspection of the autocovariance function (ACF) and partial autocovariance function (PACF). 
Model selection criteria, such as the Akaike information criterion (AIC), the corrected AIC (AICc), and the Bayesian information criterion (BIC), are also used for order selection.
 
The function \code{auto.arima()} in package \CRANpkg{forecast} provides an automatic algorithm which combines unit root tests, minimisation of the AICc in a stepwise greedy search, and MLE, to select the order of an ARIMA model. 
Here, an island parallel GAs approach is used for order selection.

\begin{figure}[htbp]
\centering\footnotesize
\begin{minipage}{0.49\textwidth}
  \centering 
  \includegraphics[width=\textwidth]{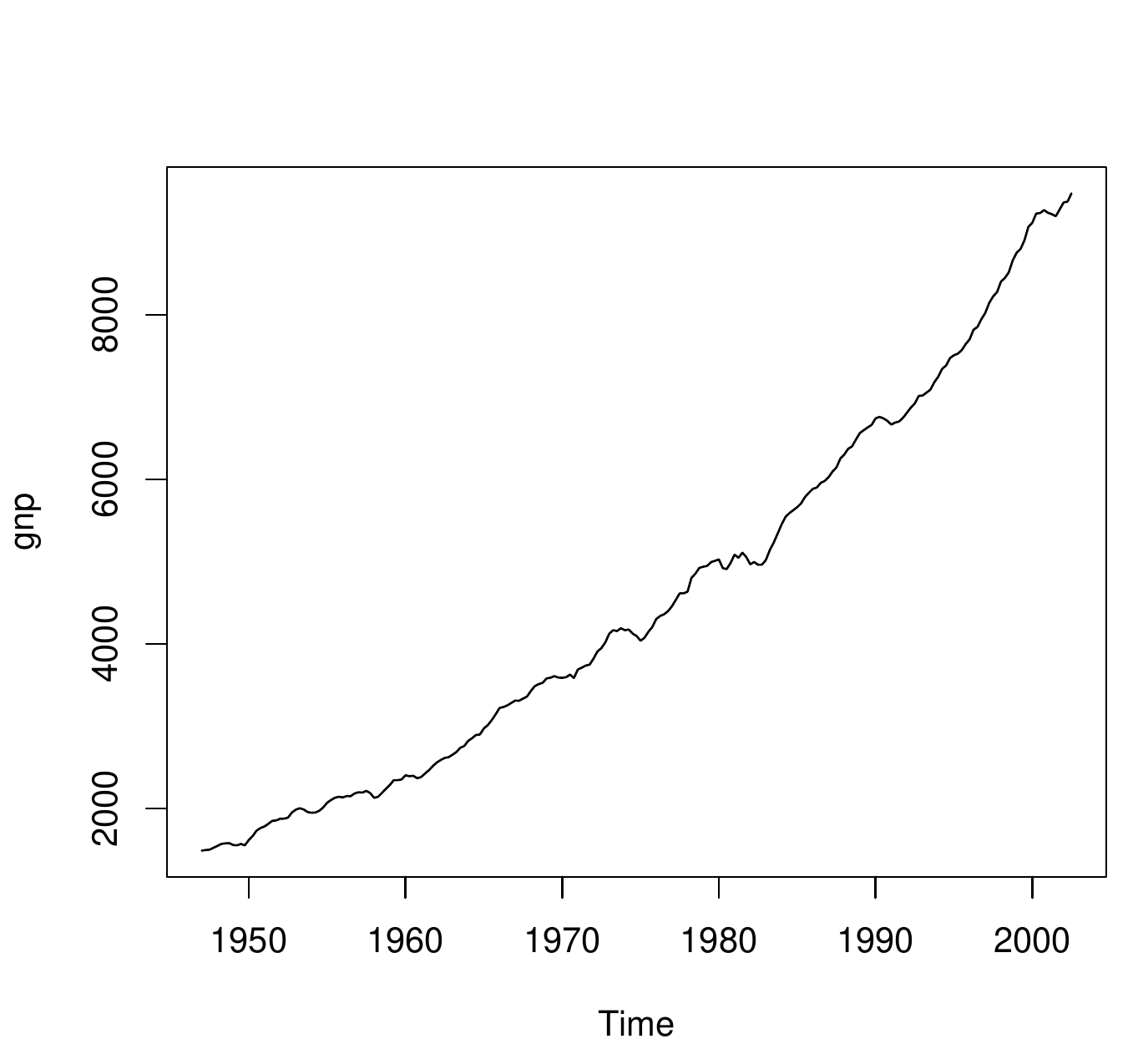}
  (a)
\end{minipage}
\begin{minipage}{0.49\textwidth}
  \centering 
  \includegraphics[width=\textwidth]{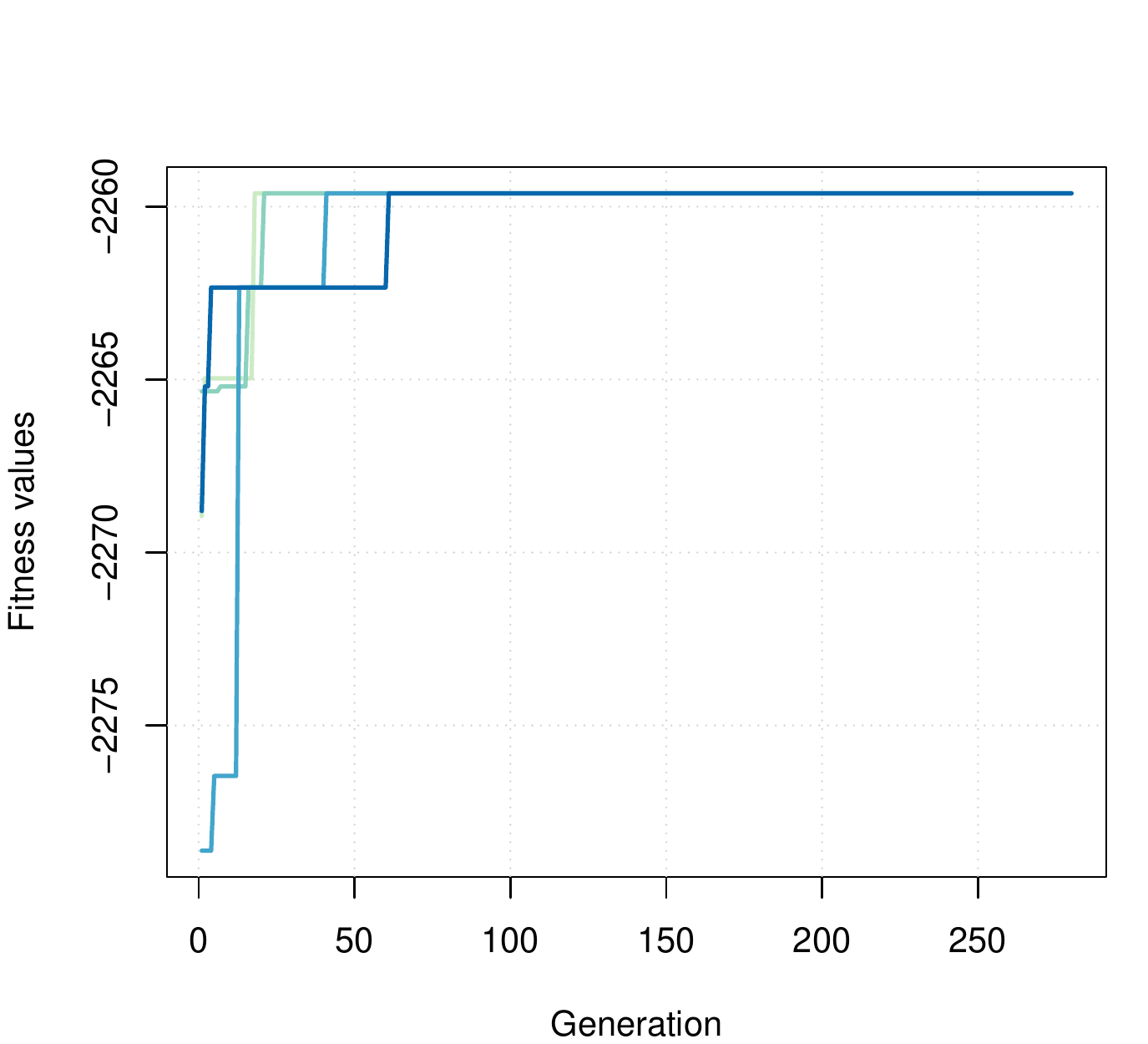}
  (b)
\end{minipage}
  \caption{(a) Plot of quarterly U.S. GNP from 1947(1) to 2002(3). (b) Trace of island parallel GAs search for ARIMA order selection.}
  \label{fig1-2:usgnp}
\end{figure}

Consider the quarterly U.S. GNP from 1947(1) to 2002(3) expressed in billions of chained 1996 dollars and seasonally adjusted. The data are available on package \CRANpkg{astsa} and described in  \citet{Shumway:Stoffer:2013}. 
\begin{example}
> data(gnp, package="astsa")
> plot(gnp)
\end{example}
The plot of the time series obtained with the last command is shown in Figure~\ref{fig1-2:usgnp}a. 

The selection of the ``optimal'' ARIMA$(p,d,q)$ model can be pursued by using binary GAs to maximise the BIC. The decision variables to be optimised are expressed in binary digits using the following function:
\begin{example}
> decode <- function(string, bitOrders)
  {
    string <- split(string, rep.int(seq.int(bitOrders), times = bitOrders)) 
    orders <- sapply(string, function(x) { binary2decimal(gray2binary(x)) })
    return(unname(orders))
  }
\end{example}
For example, using 3 bits for encoding $p$ and $q$, and 2 bits for $d$, an ARIMA(3,1,1) model can be expressed with the binary string $(0,1,0,\;0,1,\;0,0,1)$:
\begin{example}
> decode(c(0,1,0, 0,1, 0,0,1), bitOrders = c(3,2,3))
[1] 3 1 1
\end{example}
Note that the \code{decode()} function assumes that the input binary string is expressed using Gray encoding, which ensures that consecutive values have the same Hamming distance \citep{Hamming:1950}.

The fitness function to be used in the GA search is defined as follows:
\begin{example}
> fitness <- function(string, data, bitOrders)
  {
    orders <- decode(string, bitOrders)
    mod <- try(Arima(data, order = orders, include.constant = TRUE, method = "ML"),
               silent = TRUE)
    if(inherits(mod, "try-error")) NA else -mod$bic
  }
\end{example}
Note that the objective function is defined as (minus) the BIC for the specified ARIMA model, with the latter fitted using the \code{Arima()} function available in the \R\ package \pkg{forecast}.

An island binary parallel GA is then used to search for the best ARIMA model, using a migration interval of 20 generations, and the default migration rate of 0.1:
\begin{example}
> GA <- gaisl(type = "binary", nBits = 8,
              fitness = fitness, data = gnp, bitOrders = c(3,2,3),
              maxiter = 1000, run = 100, popSize = 50,
              numIslands = 4, migrationInterval = 20)
> plot(GA)
> summary(GA)
+-----------------------------------+
|         Genetic Algorithm         |
|           Islands Model           |
+-----------------------------------+

GA settings: 
Type                  =  binary 
Number of islands     =  4 
Islands pop. size     =  12 
Migration rate        =  0.1 
Migration interval    =  20 
Elitism               =  1 
Crossover probability =  0.8 
Mutation probability  =  0.1 

GA results: 
Iterations              = 280 
Epochs                  = 14
Fitness function values = -2259.615 -2259.615 -2259.615 -2259.615 
Solutions = 
     x1 x2 x3 x4 x5 x6 x7 x8
[1,]  0  1  1  1  1  0  0  1
[2,]  0  1  1  1  1  0  0  1
[3,]  0  1  1  1  1  0  0  1
[4,]  0  1  1  1  1  0  0  1
\end{example}
Figure~\ref{fig1-2:usgnp}b shows the trace of the ISLPGA search for each of the four islands used. All the islands converge to the same final solution, as also shown by the summary output above.
The selected model is an ARIMA(2,2,1), which can be fitted using:
\begin{example}
> (orders <- decode(GA@solution[1,], c(3,2,3)))
[1] 2 2 1
> mod <- Arima(gnp, order = orders, include.constant = TRUE, method = "ML")
> mod
Series: gnp 
ARIMA(2,2,1)                    

Coefficients:
         ar1     ar2      ma1
      0.2799  0.1592  -0.9735
s.e.  0.0682  0.0682   0.0143

sigma^2 estimated as 1451:  log likelihood=-1119.01
AIC=2246.02   AICc=2246.21   BIC=2259.62
\end{example}
It is interesting to compare the above solution with that obtained with the automatic procedure implemented in \code{auto.arima()} using the same criterion:
\begin{example}
> mod1 <- auto.arima(gnp, ic = "bic")
> print(mod1)
Series: gnp 
ARIMA(1,2,1)                    

Coefficients:
         ar1      ma1
      0.3243  -0.9671
s.e.  0.0665   0.0162

sigma^2 estimated as 1486:  log likelihood=-1121.71
AIC=2249.43   AICc=2249.54   BIC=2259.62
> mod1$bic
[1] 2259.622
> mod$bic
[1] 2259.615
\end{example}
The model returned by \code{auto.arima()} is an ARIMA(1,2,1), so a simpler model where an AR(1) component is chosen instead of an AR(2). The BIC values are almost equivalent, with a slightly smaller value for the ARIMA(2,2,1) model identified by ISLPGA. However, by looking at some diagnostic plots it seems that a second-order AR component is really needed to account for autocorrelation at several lags as indicated by the Ljung-Box test of autocorrelation (see Figure~\ref{fig3:usgnp}; the code used to produce the plots is available in the supplementary material).

\begin{figure}[htbp]
  \centering
  \includegraphics[width=0.9\textwidth]{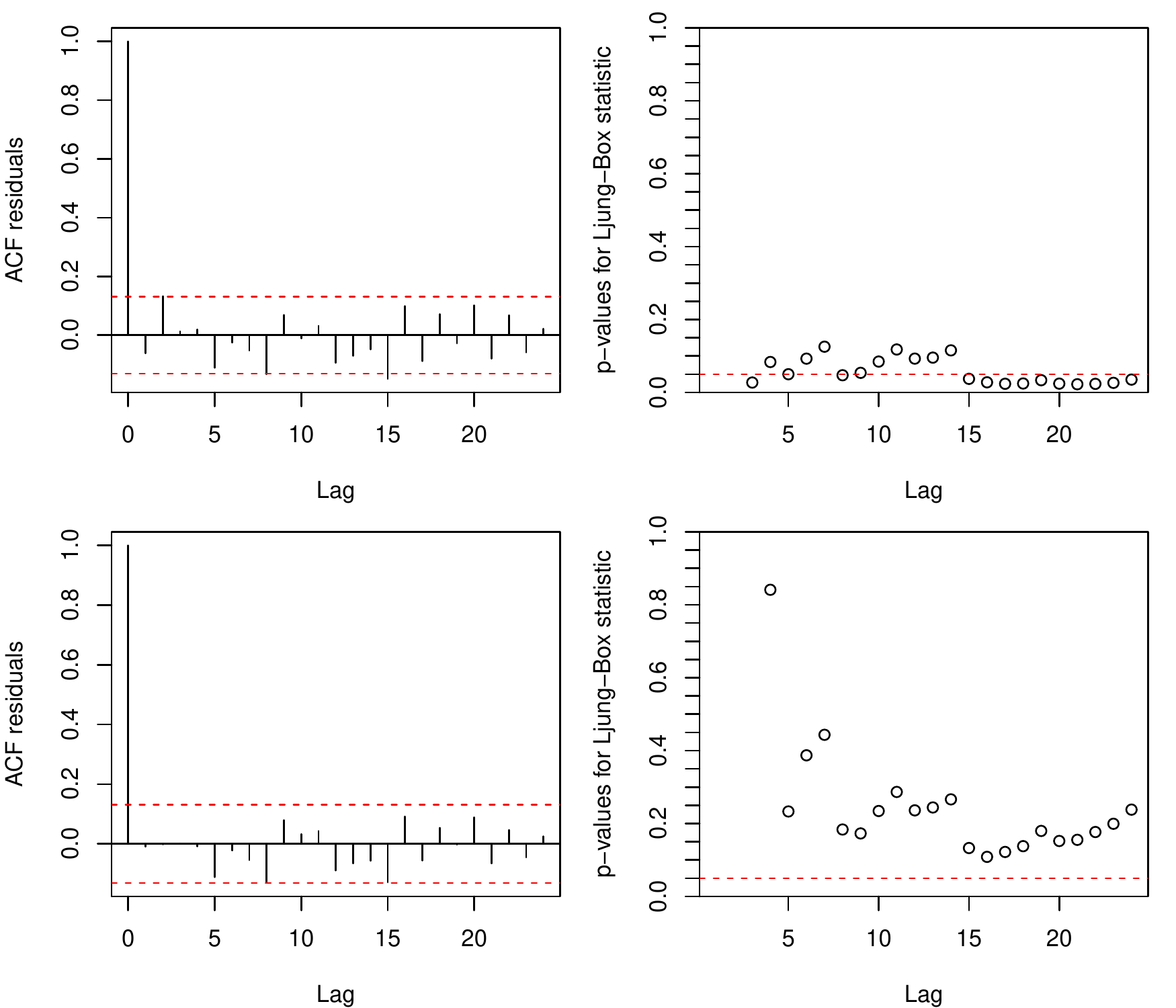}
  \caption{ACF of residuals and p-values for the Ljung-Box test of autocorrelation for the ARIMA(1,2,1) model (top graphs) and the ARIMA(2,2,1) model (bottom graphs) fitted to the quarterly U.S. GNP data from 1947(1) to 2002(3).}
  \label{fig3:usgnp}
\end{figure}

\subsection{Empirical Bayes beta-binomial model for rates estimation}

Consider the problem of estimating the probability $p_i$ of an event based on the observed number of successes $x_i$ out of $n_i$ trials, for $i=1,\ldots,N$ independent observations.
A reasonable model assumes a binomial distribution for the number of successes, i.e.  
$$
x_i|p_i \sim Bin(p_i, n_i),
$$ 
with known trials $n_i > 0$ and unknown parameters $p_i$. 
Suppose that the $p_i$s are generated from a common distribution, which we may take to be the Beta distribution, i.e. 
$$
p_i \sim Be(\alpha, \beta).
$$ 
This a conjugate prior for the binomial likelihood, so the posterior distribution turns out to be 
$$
p_i|x_i \sim Be(\alpha + x_i, \beta + n_i - x_i).
$$
The unknown rate $p_i$ can then be estimated by the posterior mean 
$E(p_i|x_i) = \dfrac{\alpha + x_i}{\alpha + \beta + n_i}$,
or by the maximum a posteriori estimate, 
$MAP(p_i|x_i) = \dfrac{\alpha + x_i - 1}{\alpha + \beta + n_i - 2}$.

In the Empirical Bayes approach the parameters $\alpha$ and $\beta$ of the prior distribution are estimated using the observed data. This is done by maximising the marginal likelihood of $x$ obtained by integrating the distribution of $x_i|p_i$ with respect to the parameter $p_i$. Thus, omitting the subscript $i$, we may write
\begin{align*}
f(x|\alpha,\beta,n) & = \int_{0}^{1} Bin(x|p,n) Be(p|\alpha,\beta) dp \\
& = \int_{0}^{1} \left\{ \binom{n}{x} p^x (1-p)^{n-x}
                         \frac{p^{\alpha-1}(1-p)^{\beta-1}}{B(\alpha,\beta)} 
                 \right\} dp \\
& = \binom{n}{x} \frac{B(\alpha + x, \beta +n -x)}{B(\alpha,\beta)},
\end{align*}
where $B(\alpha,\beta)=\Gamma(\alpha)\,\Gamma(\beta)/\Gamma(\alpha+\beta)$ is the beta function, with $\Gamma(t) = \int_0^\infty  x^{t-1} e^{-x}\,dx$. This is the density of a Beta-Binomial distribution, for which we can write
$$
x_i \sim BeBin(\alpha, \beta, n_i)
$$
with 
$$
E(x_i) = n_i \frac{\alpha}{\alpha+\beta},
$$
and
$$
Var(x_i) = n_i \frac{\alpha\beta}{(\alpha+\beta)^2}
\frac{\alpha+\beta+n_i}{\alpha+\beta+1}.
$$
 
Under the independence assumption, the marginal log-likelihood is then
\begin{equation}
\ell(\alpha,\beta) = 
 \sum_{i=1}^{n} \left\{ 
   \log\binom{n_i}{x_i} + \log B(\alpha + x_i, \beta + n_i - x_i) 
   - \log B(\alpha, \beta)
 \right\}.
\label{eq:mloglik_betabinom}
\end{equation}
 
In the Empirical Bayes approach the general idea is to estimate the parameters of the prior distribution from the data, rather than fixing them based on prior knowledge. Thus, the MMLE of parameters $(\alpha,\beta)$ are obtained by maximising the marginal log-likelihood in \eqref{eq:mloglik_betabinom}, which are then used to obtain the posterior distribution.

\begin{figure}[htbp]
\centering\footnotesize
\begin{minipage}{0.49\textwidth}
  \centering 
  \includegraphics[width=\textwidth]{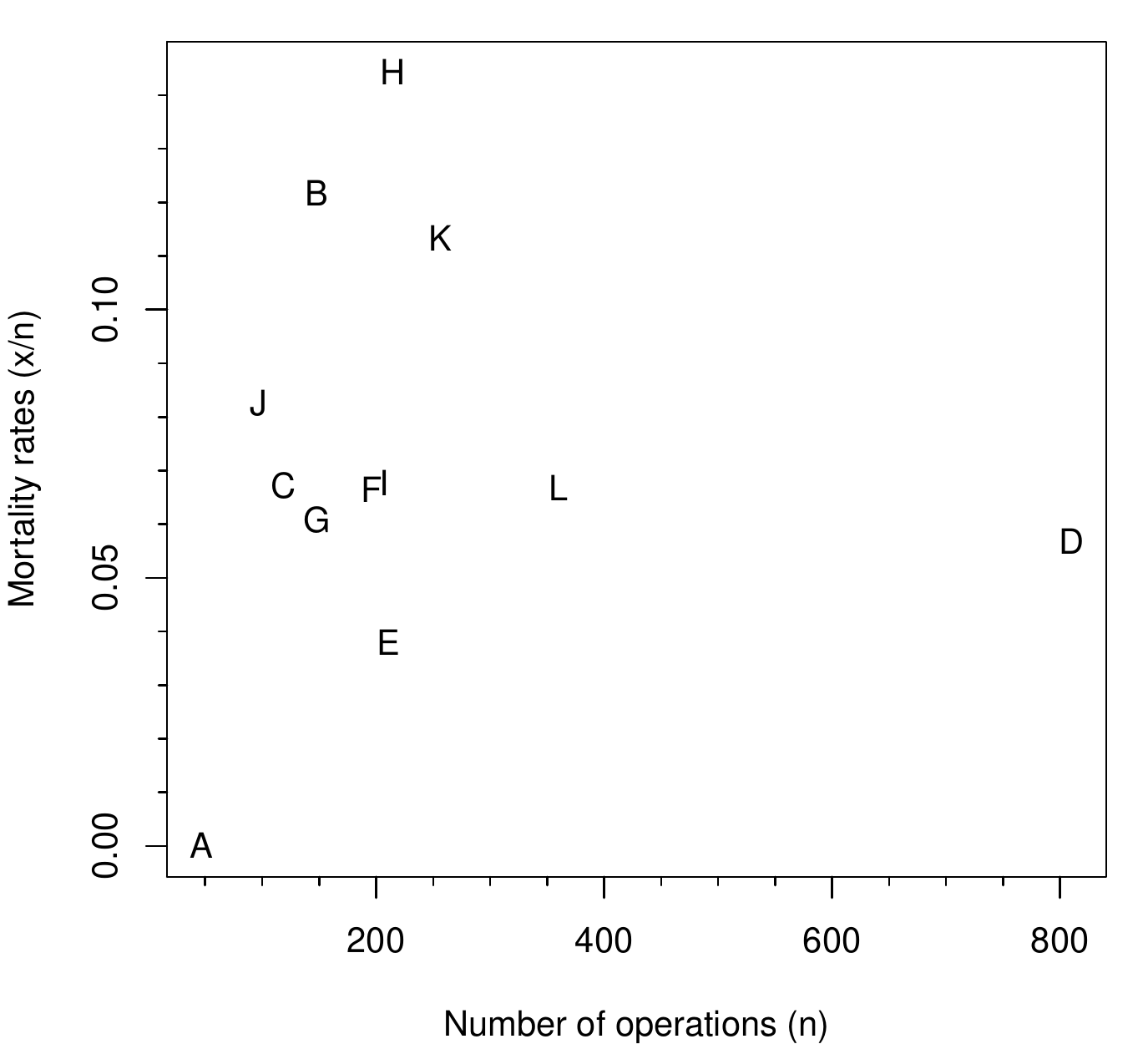}
  (a)
\end{minipage}
\begin{minipage}{0.49\textwidth}
  \centering 
  \includegraphics[width=\textwidth]{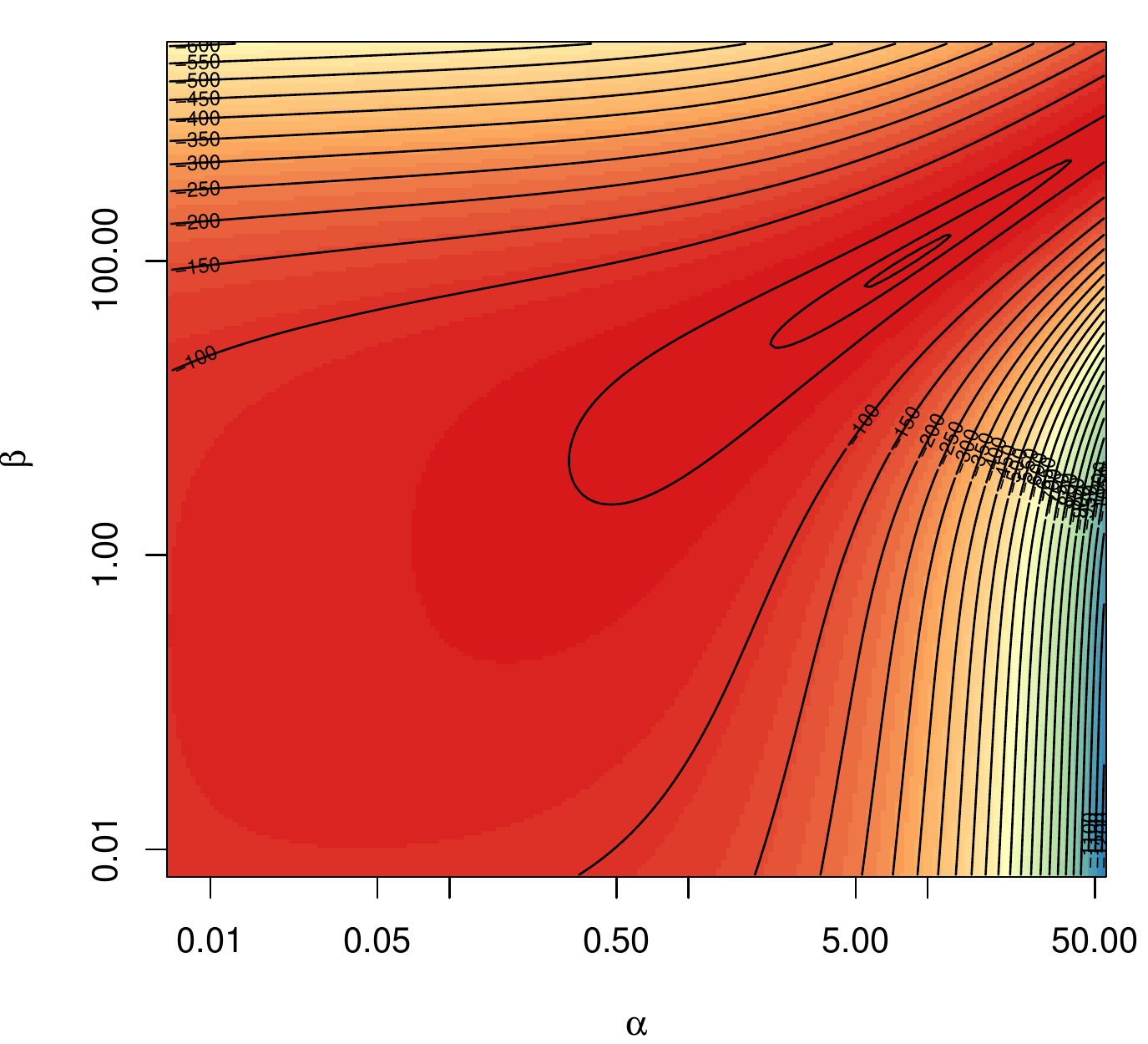}
  (b)
\end{minipage}
  \caption{(a) Plot of mortality rates for cardiac surgery on babies at 12 Hospitals. (b) Contour plot of the marginal log-likelihood surface with axes for the parameters on the log scale.}
  \label{fig12:surgical}
\end{figure}

Consider the data on mortality rates in 12 hospitals performing cardiac surgery on babies \citep[p. 15]{Spiegelhalter:etal:1996} and available in the \R\ package \CRANpkg{SMPracticals}.
\begin{example}
> data("cardiac", package = "SMPracticals")
> x <- cardiac$r
> n <- cardiac$m
> Hospitals <- rownames(cardiac)
> plot(n, x/n, type = "n",
       xlab = "Number of operations (n)", 
       ylab = "Mortality rates (x/n)")
> text(n, x/n, Hospitals)
\end{example}
As shown in Figure~\ref{fig12:surgical}a there exists a large variability on the number of operations $n_i$, ranging from Hospital A with 47 cardiac operations to Hospital D with more than 800. The ratios $x_i/n_i$ are the MLE for the mortality rates, but they strongly depend on the number of surgeries performed. For example, the mortality rate of 0 for the Hospital A is likely the result of the limited number of operations carried out.

The marginal log-likelihood in \eqref{eq:mloglik_betabinom} can be written as
\begin{example}
> mloglik <- function(par, x, size)
  {
    a <- par[1]
    b <- par[2]
    sum(lchoose(size, x) + lbeta(a+x, b+size-x) - lbeta(a, b))
  }
\end{example}
A plot of the log-likelihood surface is shown in Figure~\ref{fig12:surgical}b and can be obtained using the following code:
\begin{example}
> ngrid <- 200
> a <- exp(seq(-5, 4, length.out = ngrid))
> b <- exp(seq(-5, 8, length.out = ngrid))
> grid <- expand.grid(a, b)
> mll <- function(par) mloglik(par, x, n)
> l <- matrix(apply(grid, 1, mll), ngrid, ngrid)
> image(a, b, l, col = spectral.colors(51), log = "xy", 
        xlab = expression(alpha), ylab = expression(beta), axes = FALSE)
> axis(1); axis(2); box()
> contour(a, b, l, nlevels = 21, add = TRUE)
> contour(a, b, l, levels = quantile(l,c(0.99,0.999)), drawlabels = FALSE, add = TRUE)
\end{example}

We opted to use parallel GAs evolving in four islands with the default immigration policies, using also a local optimisation search to speed up convergence to the optimal solution. 
\begin{example}
> GA <- gaisl("real-valued", 
              fitness = mloglik, x = x, size = n,
              min = exp(c(-5,-5)), max = exp(c(4,8)), names = c("a", "b"),
              numIslands = 4, optim = TRUE, 
              maxiter = 1000, run = 200)
> plot(GA, log = "x")
> summary(GA)
+-----------------------------------+
|         Genetic Algorithm         |
|           Islands Model           |
+-----------------------------------+

GA settings: 
Type                  =  real-valued 
Number of islands     =  4 
Islands pop. size     =  25 
Migration rate        =  0.1 
Migration interval    =  10 
Elitism               =  1 
Crossover probability =  0.8 
Mutation probability  =  0.1 
Search domain = 
             a            b
Min  0.0067379    0.0067379
Max 54.5981500 2980.9579870

GA results: 
Iterations              = 220 
Epochs                  = 22 
Fitness function values = -38.753 -38.753 -38.753 -38.753 
Solutions = 
          a      b
[1,] 8.2535 99.637
[2,] 8.2535 99.637
[3,] 8.2535 99.637
[4,] 8.2535 99.637
\end{example}

\begin{figure}[htbp]
\begin{minipage}{0.5\textwidth}
  \centering
  \includegraphics[width=\textwidth]{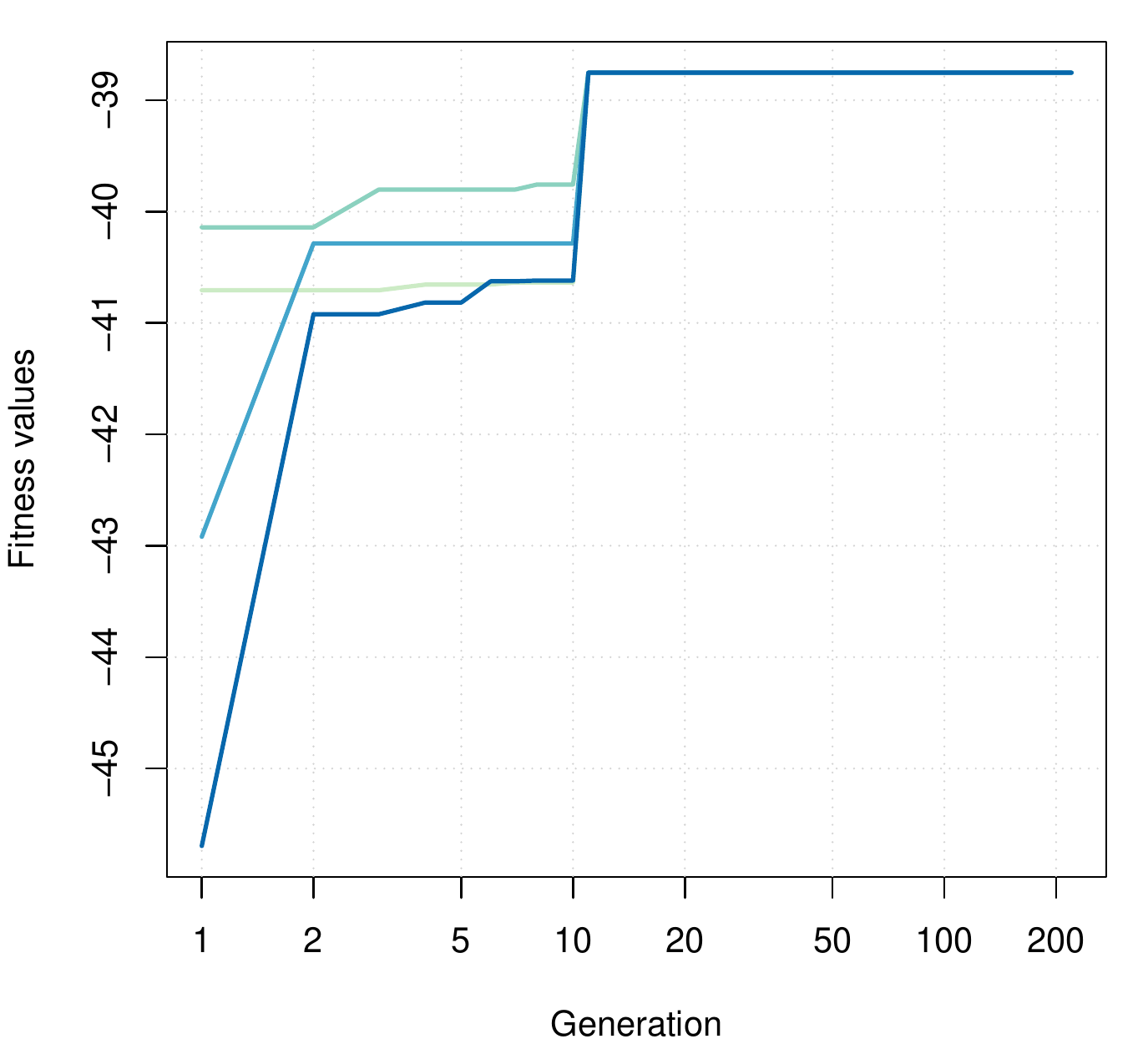}
  (a)
\end{minipage}
\begin{minipage}{0.5\textwidth}
  \centering
  \includegraphics[width=\textwidth]{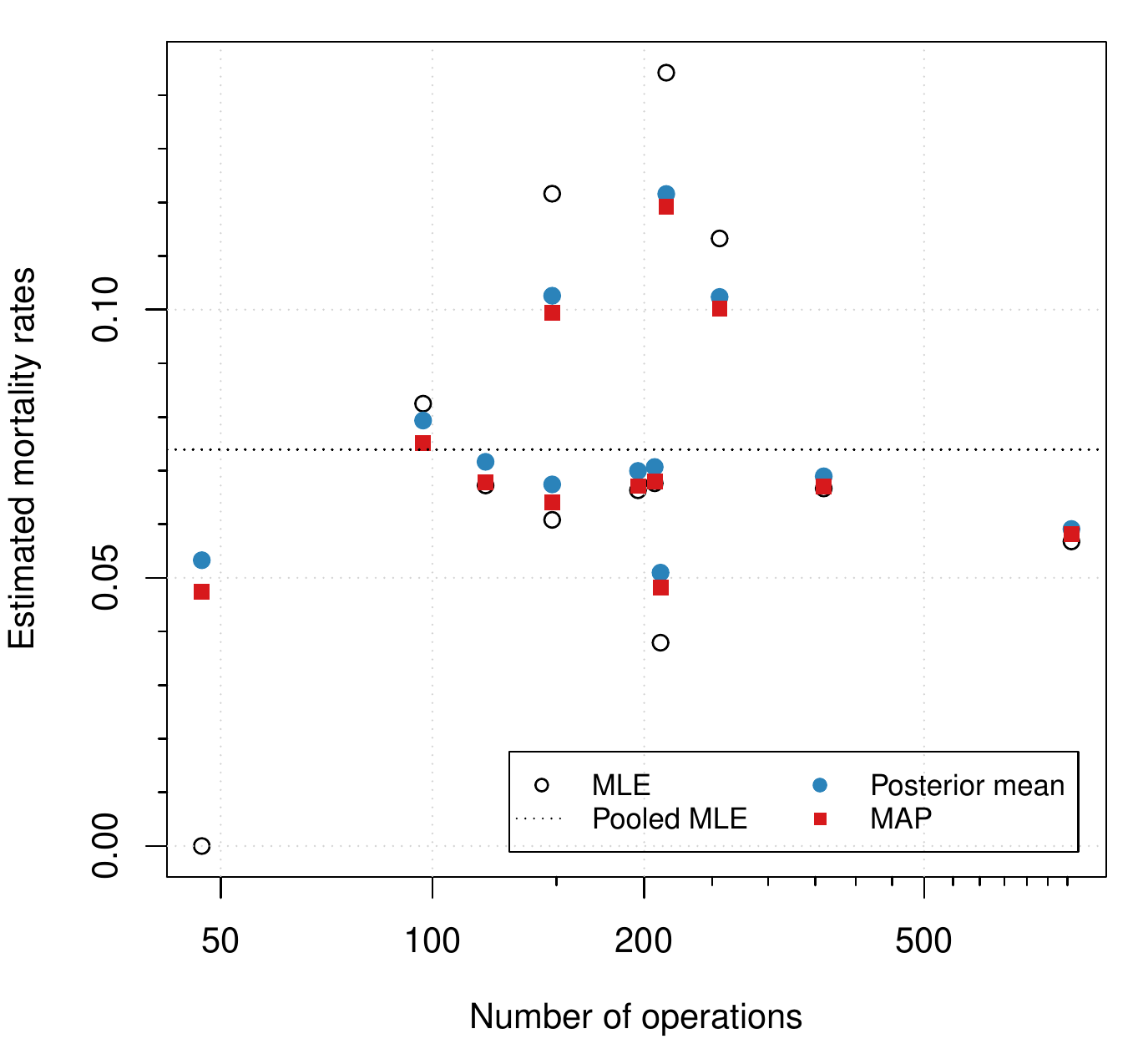}
  (b)
\end{minipage}
  \caption{(a) Trace of GA evolution in each island, with the x-axis on the log scale to enhance the first few iterations. (b) Plot of estimated mortality rates vs the number of surgical operations, with the x-axis on the log scale.}
  \label{fig34:surgical}
\end{figure}

Looking at the trace of GA evolution in each island as shown in Figure~\ref{fig34:surgical}a, we can see that the algorithm soon achieves the optimal value for all the islands and then remain constants until a stopping rule is meet. The solution found is $(\hat{\alpha} = 8.2535, \hat{\beta} = 99.637$), which can be used to compute the posterior mean and the MAP estimate. For completeness we also compute the MLE and pooled MLE values:
\begin{example}
> (MLE <- x/n)
 [1] 0.000000 0.121622 0.067227 0.056790 0.037915 0.066327 0.060811 0.144186 0.067633
[10] 0.082474 0.113281 0.066667
> (pooledMLE <- sum(x)/sum(n))
[1] 0.073916
> par <- GA@solution[1,]
> (posteriorMean <- (par[1] + x)/(par[1] + par[2] + n))
 [1] 0.053286 0.102597 0.071636 0.059107 0.050969 0.069938 0.067425 0.121569 0.070671
[10] 0.079328 0.102376 0.068934
> (MAP <- (par[1] + x - 1)/(par[1] + par[2] + n - 2))
 [1] 0.047442 0.099466 0.067826 0.058144 0.048135 0.067089 0.064018 0.119210 0.067926
[10] 0.075181 0.100178 0.067083
\end{example}
The estimates are shown graphically with the code
\begin{example}
> plot(n, MLE, log = "x", 
       xlab = "Number of operations",
       ylab = "Estimated mortality rates")
> axis(1, at = seq(50,800,by=50), tck=-0.01, labels = FALSE)
> axis(2, at = seq(0.01,0.15,by=0.01), tck=-0.01, labels = FALSE)
> grid(equilogs = FALSE)
> points(n, posteriorMean, col = spectral.colors(2)[1], pch = 19)
> points(n, MAP, col = spectral.colors(2)[2], pch = 15)
> abline(h = pooledMLE, lty = 3)
> legend("bottomright", legend = c("MLE", "Pooled MLE", "Posterior mean", "MAP"), 
         col = c(1,1,spectral.colors(2)), pch = c(1,NA,19,15), lty = c(NA,3,NA,NA),
         ncol = 2, inset = 0.03, cex = 0.8)
\end{example}
From Figure~\ref{fig34:surgical}b we can see that EB estimates for the mortality rates are shrunk toward the overall average (corresponding to the pooled MLE), with the posterior mean uniformly larger than the MAP due to the fact that the distribution is skewed to the right. EB prior estimation has a small effect on the Hospitals with larger number of surgical operations, whereas it has a large impact on those hospitals with small number of operations (e.g. Hospital A) or on those with more extreme rates (e.g. Hospitals H, B, K, and E).

\begin{figure}[htbp]
  \centering
  \includegraphics[width=0.8\textwidth]{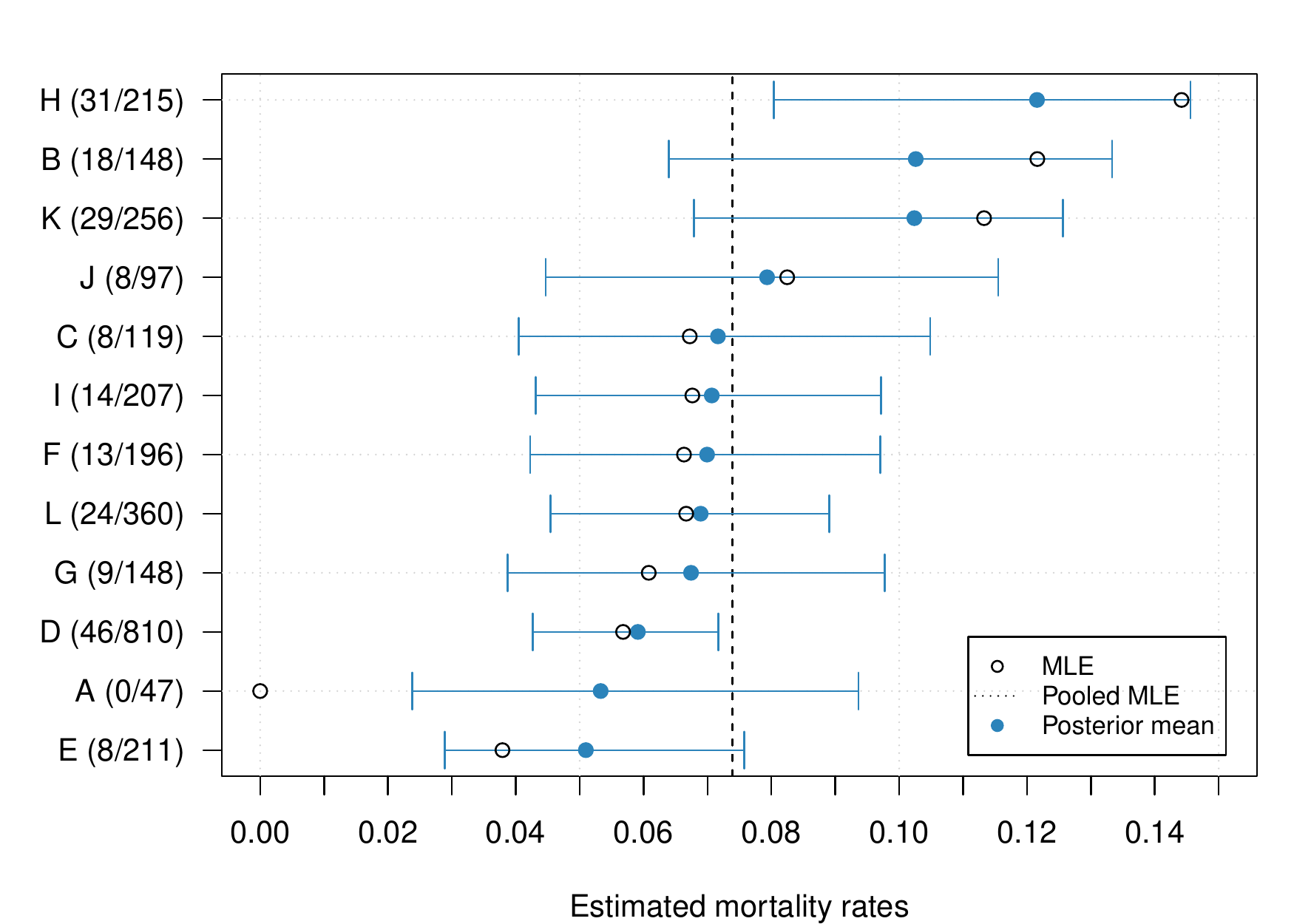}
  \caption{MLEs and posterior means with 95\% confidence intervals for the surgical mortality rates in each hospital. Numbers in brackets show the observed number of deaths and the total number of operations. The dashed vertical line indicates the population mean failure rate (pooled MLE).}
  \label{fig5:surgical}
\end{figure}

Although better approaches are available \citep[sec. 3.5]{Carlin:Louis:2000}, and in particular that proposed in \citet{Laird:Louis:1987}, equi-tail \emph{naive} Empirical Bayes confidence intervals can be easily obtained from the quantiles of the Beta distribution:
\begin{example}
> level <- 0.95
> EBconfint <- data.frame(lower = numeric(length(x)), 
                          upper = numeric(length(x)))
> for(i in 1:nrow(EBconfint))
  { 
    EBconfint[i,] <- qbeta(c((1-level)/2, (1+level)/2), 
                           shape1 = (par[1] + x[i]), 
                           shape2 = (par[2] + n[i]))
  }
> (tab <- data.frame(x, n, MLE, pooledMLE, MAP, posteriorMean, EBconfint))
    x   n      MLE pooledMLE      MAP posteriorMean    lower    upper
1   0  47 0.000000  0.073916 0.047442      0.053286 0.023805 0.093613
2  18 148 0.121622  0.073916 0.099466      0.102597 0.063953 0.133335
3   8 119 0.067227  0.073916 0.067826      0.071636 0.040438 0.104879
4  46 810 0.056790  0.073916 0.058144      0.059107 0.042638 0.071677
5   8 211 0.037915  0.073916 0.048135      0.050969 0.028915 0.075727
6  13 196 0.066327  0.073916 0.067089      0.069938 0.042261 0.097039
7   9 148 0.060811  0.073916 0.064018      0.067425 0.038717 0.097751
8  31 215 0.144186  0.073916 0.119210      0.121569 0.080397 0.145603
9  14 207 0.067633  0.073916 0.067926      0.070671 0.043145 0.097148
10  8  97 0.082474  0.073916 0.075181      0.079328 0.044698 0.115510
11 29 256 0.113281  0.073916 0.100178      0.102376 0.067903 0.125637
12 24 360 0.066667  0.073916 0.067083      0.068934 0.045445 0.089036
\end{example}
and shown graphically in Figure~\ref{fig5:surgical} using the following code 
\begin{example}
> ord <- order(tab$posteriorMean)
> par(mar = c(4,6,2,1))
> with(tab[ord,],
  { plot(0, 0, ylim = range(ord), xlim = c(0,0.15), xaxt = "n", yaxt = "n",
         xlab = "Estimated mortality rates", ylab = "") 
    axis(side = 1, at = seq(0,0.15,by=0.01))
    axis(side = 2, at = seq(ord), las = 2,
         labels = paste0(rownames(cardiac)[ord], " (",  x, "/", n, ")"))
    grid()
    abline(v = pooledMLE, lty = 2)
    mclust:::errorBars(seq(ord), lower, upper, col = spectral.colors(2)[1], horizontal = TRUE)
    points(posteriorMean, seq(ord), pch = 19, col = spectral.colors(2)[1])
    points(MLE, seq(ord), pch = 1)
  })
\end{example}

\subsection{Benchmark function optimisation}
\label{sec:benchopt}

\citet{Mullen:2014} compared several optimisation algorithms using 48 benchmark functions available in the \CRANpkg{globalOptTests} package. \pkg{GA} was one of the several \R\ packages investigated in such a comparison. However, with the settings used in this study, its overall performance was not particularly brilliant, ranking 14th out of 18 methods, thus leaving plenty of room for improvements.

\begin{figure}[htbp]
\centering\footnotesize
\begin{minipage}{0.49\textwidth}
  \centering
  \includegraphics[width=\textwidth]{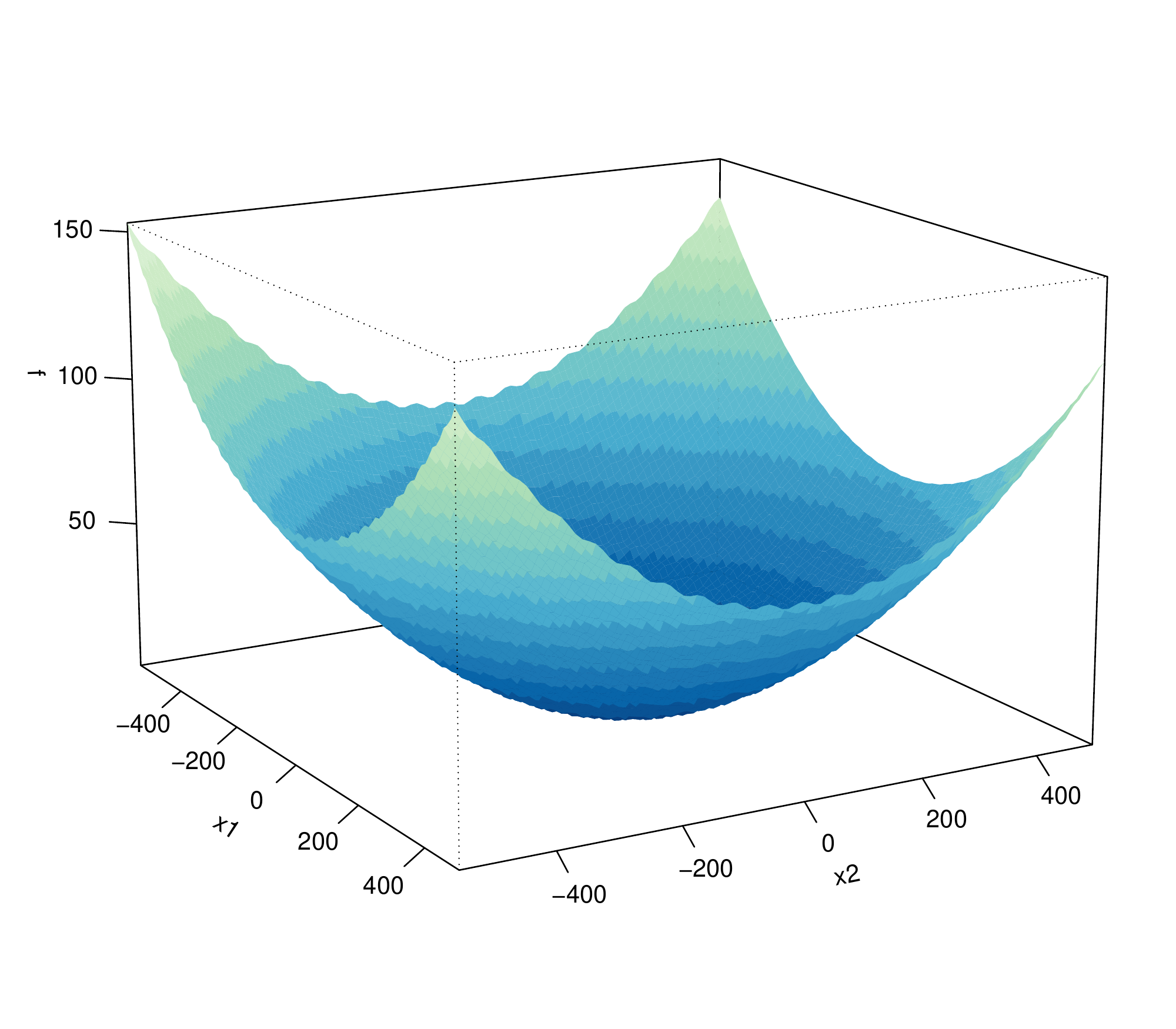}
  (a)
\end{minipage}
\begin{minipage}{0.49\textwidth}
  \centering
  \includegraphics[width=\textwidth]{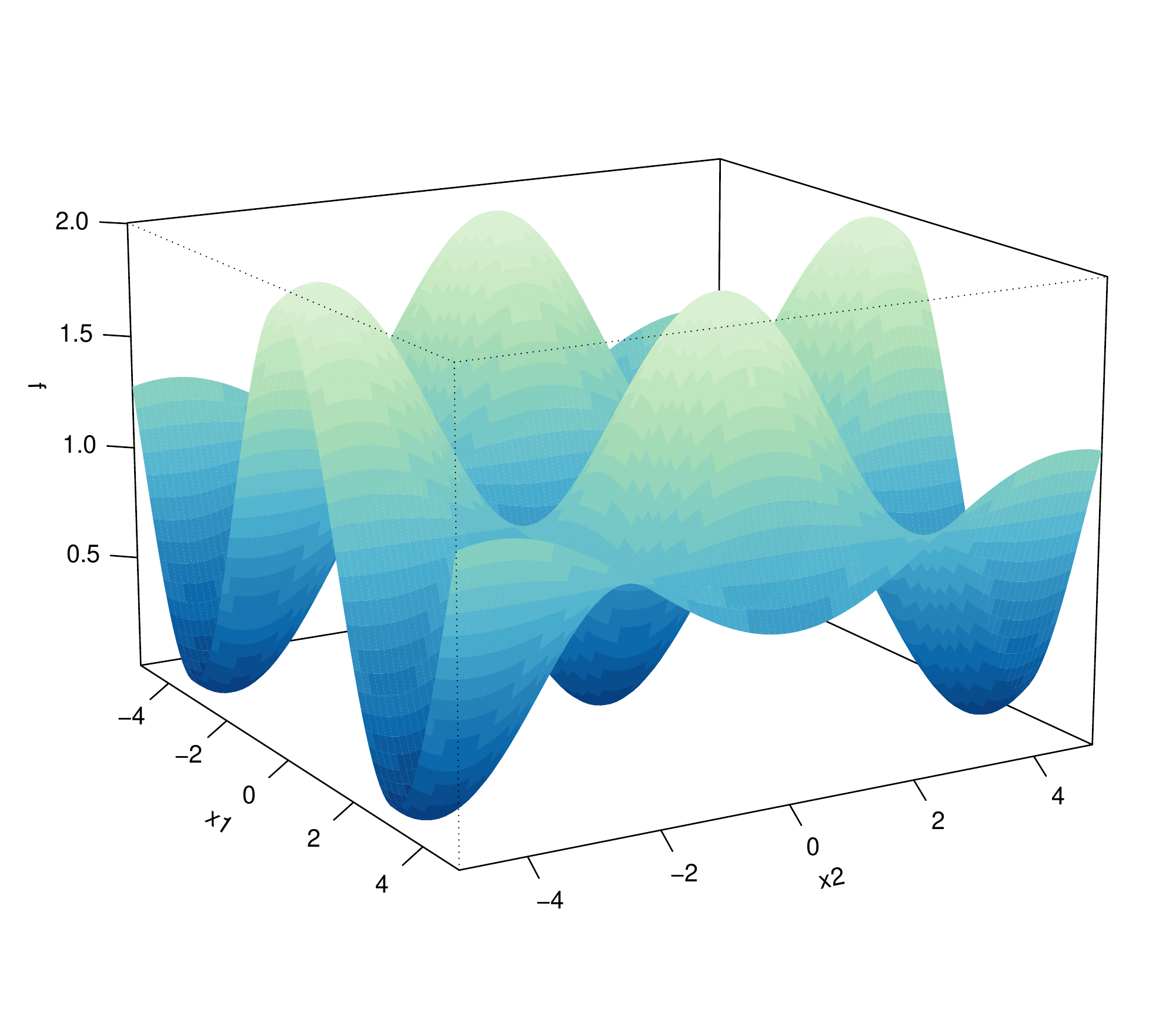}
  (b)
\end{minipage}
  \caption{Perspective plots of two-dimensional Griewank function. At larger scale the function appears convex (a), but zooming reveals a complex structure with numerous local minima (b).}
  \label{fig1:Griewank}
\end{figure}

One of the problematic case is the Griewank function, which is defined as
$$
f(x_1, \ldots, x_d) = 1 + \frac{1}{4000} \sum_{i=1}^d x^2_i - \prod_{i=1}^d \cos(x_i/\sqrt{i}).
$$ 
This a multimodal, non-separable function, with several local optima within the search region. 
For any dimensionality $d$, it has one global minimum of zero located at the point $(0, \ldots, 0)$. Figure~\ref{fig1:Griewank} shows some perspective plots for $d=2$ at different zooming levels.

\begin{figure}[htbp]
\centering\footnotesize
\includegraphics[width=0.7\textwidth]{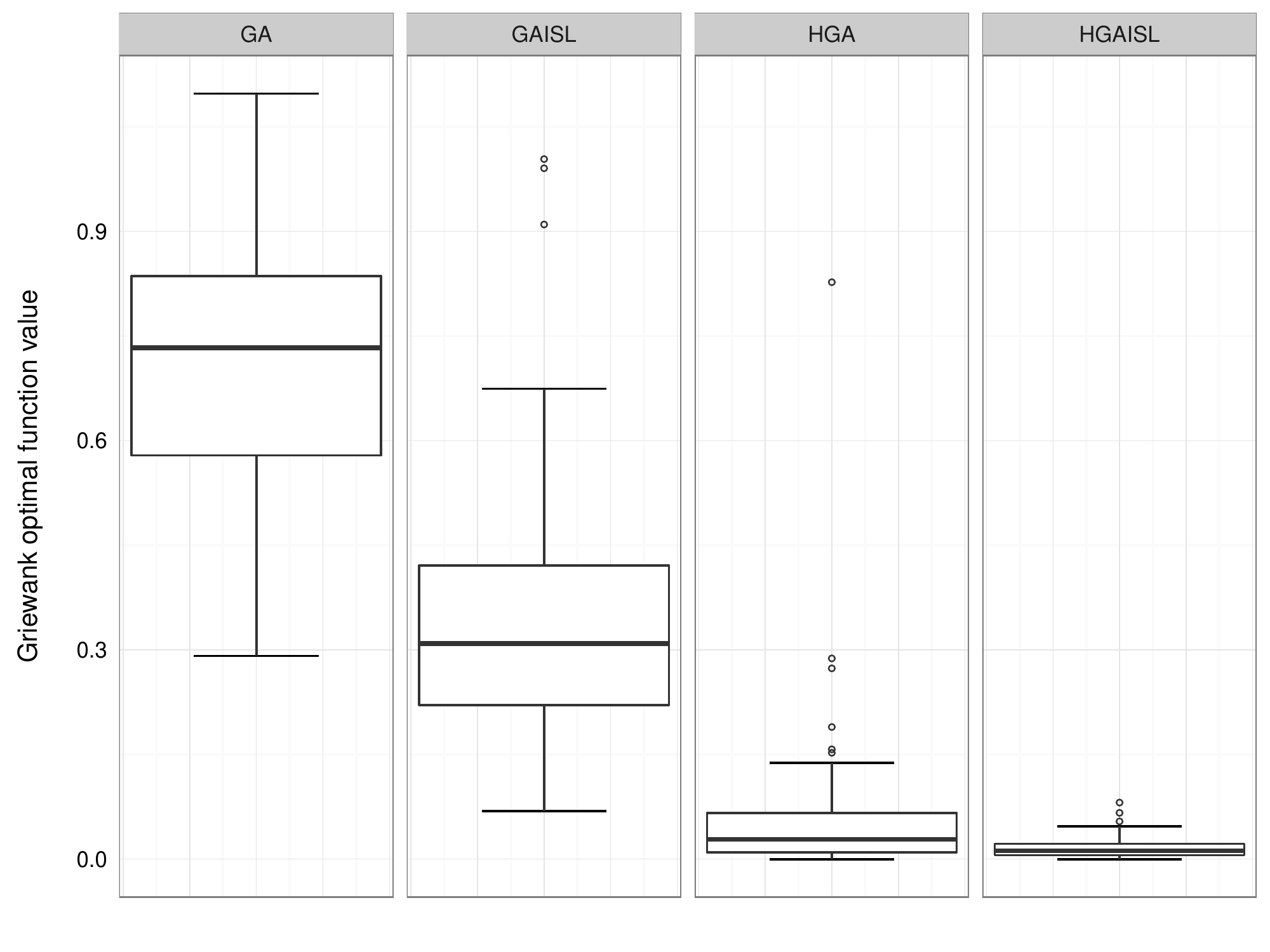}
\caption{Results from 100 replications of Griewank function optimisation using standard GAs (\code{GA}), island GAs (\code{GAISL}), hybrid GAs with local search (\code{HGA}), and island GAs with local search (\code{HGAISL}).}
\label{fig1:GriewankBenchmark}
\end{figure}

We replicated the simulation study in \citet{Mullen:2014} using the standard sequential GA (\code{GA}), the parallel island GA with 4 islands (\code{GAISL}), the hybrid GA with local search (\code{HGA}), and the island GA with local search (\code{HGAISL}). 
Results for the Griewank function based on 100 replications are shown in Figure~\ref{fig1:GriewankBenchmark}. 
The use of hybrid GAs, particularly in combination with the islands evolution, clearly yields more accurate solutions and with less dispersion. 
The same behavior has been observed in many other benchmark functions available in the \pkg{globalOptTests} package.

\citet[Section 5]{Mullen:2014} also provided a measure of accuracy for each optimisation method considered by counting the number of successful runs, with the latter defined as a solution less than $0.005$ from the minimum of the objective function. The empirical accuracy scores obtained in our simulations are shown in Table~\ref{tab1:benchopt}, and these can be compared with those reported in Mullen's paper and its supplemental material. Hybrid GAs including local optimisation search (HGA) yield a large improvement on accuracy (ranking 2nd with a score of 3717), and when combined with island evolution (HGAISL) achieve the highest overall score (ranking 1st with a score equal to 3954).

\begin{table}[!htb]
\centering
\caption{Benchmark functions accuracy scores for GAs and some hybrid and islands evolution variants (larger values are better).}
\label{tab1:benchopt}
\begin{tabular}{ccccc}
\toprule
GA from Mullen's paper  & GA & GAISL & HGA & HGAISL \\
\midrule
  2259 & 2372 & 2587 & 3717 & 3954 \\
\bottomrule
\end{tabular}
\end{table}

\section{Summary}

\pkg{GA} is a flexible \R\ package for solving optimisation problems with genetic algorithms. This paper discusses some improvements recently added to the package.
We have discussed the implementation of hybrid GAs, which employ local searches during the evolution of a GA to improve accuracy and efficiency.
Further speedup can also be achieved by parallel computing. This has been implemented following two different approaches. In the first one, the so-called master-slave approach, the fitness function is evaluated in parallel, either on a single multi-cores machine or on a cluster of multiple computers. In the second approach, called islands model, the evolution takes place independently on several sub-populations assigned to different islands, with occasional migration of solutions between islands. 
Both enhancements often lead to high-quality solutions more efficiently.

Future plans include the possibility to improve overall performance by rewriting some key functions in C++ using the \CRANpkg{Rcpp} package. In particular, coding of genetic operators in C++ should provide sensible benefits in terms of computational speedup.  
Finally, the package \CRANpkg{memoise} enables to store the results of an expensive fitness function call and returns the cached result when the same input arguments occur again. This strategy could be conveniently employed in the case of binary GAs.


\section*{Acknowledgements}

The author acknowledge the CINECA award under the ISCRA initiative (\url{http://www.hpc.cineca.it/services/iscra}) for the availability of high performance computing resources and support.

\baselineskip=14pt
\bibliographystyle{chicago}
\bibliography{GAv3_arxiv}

\end{document}